\documentclass[aps,pra,amssymb,amsmath,nobibnotes,nobibnotes,reprint,superscriptaddress,showpacs,showkeys]{revtex4-1}
\usepackage[colorlinks, linkcolor=blue, anchorcolor=blue, citecolor=blue]{hyperref}
\usepackage{graphicx}
\usepackage{revsymb4-1}
\usepackage{bm}
\usepackage{braket}
\usepackage{graphics}
\usepackage{epstopdf}
\usepackage{graphicx,graphics,color,epsfig}

\begin{document}

\title{Evidence of indistinguishability and entanglement determined by the energy-time uncertainty principle in a system of two strongly coupled bosonic modes}

\author{Smail Bougouffa}
\email{sbougouffa@hotmail.com}
\affiliation{Department of Physics, Faculty of Science, Taibah University, P.O. Box 30002, Madinah~41481, Saudi Arabia}
\author{Zbigniew Ficek}
%\email{zficek@kacst.edu.sa}
\affiliation{National Center for Applied Physics, KACST, P.O. Box 6086, Riyadh~11442, Saudi Arabia}

\date{\today}

\begin{abstract}
The link of two concepts, indistinguishability and entanglement, with the energy-time uncertainty principle is demonstrated in a system composed of two strongly coupled bosonic modes. Working in the limit of a short interaction time, we find that the inclusion of the antiresonant terms to the coupling Hamiltonian leads the system to relax to a state which is not the ground state of the system. This effect occurs passively by just presence of the antiresonant terms and is explained in terms of the time-energy uncertainty principle for the simple reason that at a very short interaction time, the uncertainty in the energy is of order of the energy of a single excitation, thereby leading to a distribution of the population among the zero, singly and doubly excited states. The population distribution, correlations and entanglement are shown to be substantially depend on whether the modes decay independently or collectively to an exterior reservoir. In particular, when the modes decay independently with equal rates, entanglement with the complete distinguishability of the modes is observed. The modes can be made mutually coherent if they decay with unequal rates. However, the visibility in the single-photon interference cannot exceed $50\%$. When the modes experience collective damping, they are indistinguishable even if decay with equal rates and the visibility can, in principle, be as large as unity. We find that this feature derives from the decay of the system to a pure entangled state rather than the expected mixed state. When the modes decay with equal rates, the steady-state values of the density matrix elements are found dependent on their initial values.
\end{abstract}

\maketitle

\section{Introduction}\label{sec1}

There has recently been a great interest in the realization of quantum networks of coupled qubits formed by spatially periodic structures of trapped atoms~\cite{st02,hm13,gh15}, arrays of coupled optical cavities~\cite{hb06,as07,zd08,tf10,kb12,sl12,m13} or superconducting electrical circuits~\cite{lf13,fg11,ma14,ht12}. Quantum networks provide an experimental platform for spatial transport of quantum states required for quantum cryptography, quantum teleportation, simulation of many-body systems, quantum information processing and quantum computation. Optical cavities are ideally suited for the implementation of quantum networks where the inter-cavity coupling might be realized through the output cavity fields which could be focused and transmitted by optical elements, for example, short optical fibers. The primary objective is to achieve strong and lossless couplings. Therefore, different coupling schemes have been proposed to accomplish an efficient transfer of photons between adjacent cavities including overlapping evanescent field modes, optical fibers or waveguides, and hopping fields, the tunneling of photons between cavities~\cite{bq14}. Exchange of information between the cavities is often affected by dissipation and decoherence induced by the unavoidable coupling to the environment. For coupling via fibers or waveguides, major obstacles are losses inside the fiber or waveguide material.

A number of theoretical and experiments studies were carried out on the simplest quantum network composed of only two cavities, and several schemes have been proposed in which an efficient transmission between the cavities could be achieved~\cite{ZOR00,HBP08,MC15}. In most treatments the cavities contained two-level atoms, and the creation of entanglement between the atoms and its transfer to the cavity modes was considered~\cite{cz97,tp97,ek99,mb04,io08,bek08,ye10,zl10,bb10,sy11,xf12,bf13}. It has also been demonstrated that effective quantum gates between atoms located in distant cavities can be realized even in the presence of losses and imperfections in coupling strengths~\cite{sm06,bk08}. In addition, the interaction of the cavities with an injected squeezed field or with a squeezed reservoir has been studied~\cite{kc04,ti10,ab15}.

The previous work on quantum networks of coupled cavities was limited to the weak coupling regime described by the coupling Hamiltonian containing only resonant terms, the photon hopping between the modes. In general, the coupling Hamiltonian also contains antiresonant terms such that the creation of an excitation in a given mode is accompanied by the creation of a negative energy quantum in the other mode. In the weak coupling regime the antiresonant terms make much smaller contributions and therefore are often omitted, under the rotating-wave approximation~\cite{ae75}. However, in the strong coupling regime in which the magnitude of the coupling strength is comparable to the frequency of the modes, the antiresonant terms make notable contributions leading to novel features~\cite{bs40,wh97,bk00,nb08,jl09,zr09,cy10,lz12,yl13}.

In this paper, we consider a pair of coupled bosonic modes represented by two single-mode cavities coupled by a short waveguide. In studying the interaction between the cavities, we include both resonant and antiresonant terms in the interaction Hamiltonian. To say this another way, we permit for two types of the interactions, linear and non-linear to contribute simultaneously to the coupling between the cavities. Notice that the inclusion of the antiresonant terms is equivalent to take into account the energy non-conserving terms in the interaction between the cavities. These terms are known to produce virtual photons which can survive only for a time $\Delta t\sim 1/\omega$, where $\omega$ is the frequency of the modes. According to the energy-time uncertainty principle, at such short times the virtual photons fail to conserve energy by an amount $\Delta E$, which is of order of the energy of a single excitation, $\Delta E\sim \hbar\omega$. This fact can lead to a redistribution of the population among states differing in energy by $\hbar\omega$. Of particular interest is the stationary limit the system attains over this short time. This requires a strong coupling of the modes and a fast damping of the modes if one would like to achieve a stationary state over such a short time. Therefore, our results apply to a short observation time and the ultra-strong coupling regime. Some results are also presented for the so-called deep strong coupling regime, corresponding the coupling strengths larger than the field frequency~\cite{cr10,br12,dl14,sz14}.

We show that the system exhibits features, in particular coherence and entanglement features that are not present in the weak coupling regime. Two cases are studied: (i) the modes decay independently, and (ii) the modes decay collectively to an external reservoir. We find that the modes decaying independently with equal rates can be found entangled and simultaneously behaving as mutually incoherent. We calculate the visibility of the interference fringes and show how the "which-path" information is made possible when the modes decay with equal rates. The "which-way" information, however, is not possible when the modes decay with unequal rates, so a mutual coherence can be established resulting in single-photon interference between the modes. We find an upper bound that the visibility cannot exceed $50\%$ when the modes decay independently.

The modes can, however, be made entangled and simultaneously exhibiting quantum interference with $100\%$ visibility if they decay collectively. We find that in this case, the modes are always indistinguishable independent of whether they decay with equal or unequal rates. In addition, we find that the collective damping can lead to the steady-state values of the density matrix which depends on initial conditions.

The paper is organized as follows. In Sec.~\ref{sec2} we introduce the model and formulate the master equation for the density operator of the system.
The equations of motion for the density matrix elements and their steady-state solutions are given in Sec.~\ref{sec3}. The equations of motion are simple enough that we can find their steady-state values analytically. In Sec.~\ref{sec4} we discuss the problem of distinguishability between the modes induced by the energy-time uncertainty principle and methods to make the modes indistinguishable. An upper bound is imposed on the visibility of the interference fringes when the modes decay independently and it can be overtaken if the modes decay collectively. In Sec.~\ref{sec5} we examine the conditions for entanglement. Some remarks are made about the connection between the one- and two-photon coherences. Finally, in Sec.~\ref{sec6}, we summarize and conclude our results.

\section{The model and approach}\label{sec2}

We consider a pair of strongly coupled bosonic modes of equal frequencies $\omega$, labelled by the suffices $A$ and $B$. The modes are represented by the annihilation and creation operators, $\hat{a}_{j}, \hat{a}^{\dagger}_{j}\, (j=A,B)$, which satisfy the commutation relation $[\hat{a}_{i},\hat{a}^{\dagger}_{j}]=\delta_{ij}$. We assume that apart from the strong dynamical influence on each other through the direct coupling, the modes can also influence on each other through modes of the reservoir to which they are damped with rates $\gamma_{A}$ and $\gamma_{B}$, respectively. We will investigate two cases in which the modes decay independently or collectively. We will refer to these cases as the decay of the modes to either separate reservoirs or a common reservoir. In order to take into account contributions of the antiresonant (non-RWA) terms, we will require the coupling strengths and damping rates to be comparable to the frequency $\omega$. In other words, we will work in the ultra-strong coupling regime. We are interested in the steady-state characteristics of the system, in which the strong coupling processes counterbalance the decay process.
\begin{figure}[h]
\center{\includegraphics[width=0.9\columnwidth]{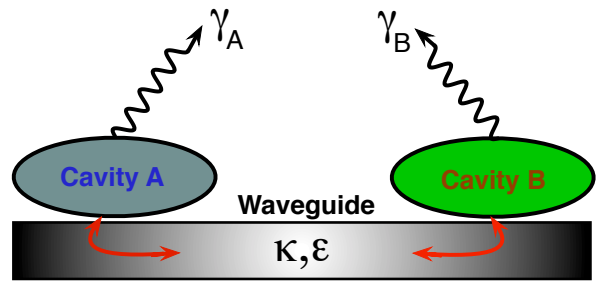}}
\caption{(Color online) Schematic diagram of the system. Two single-mode cavities are coupled to each other through a short waveguide. The photons in the cavities leak out to the waveguide with a very short leaking  time. Both resonant and antiresonant coupling processes are taken into account and are described by the coupling strengths $\kappa$ and $\epsilon$. The cavities also decay, separately or collectively, to the external environment.}
\label{fig1}
\end{figure}

In practice this model could be realized in a circuit QED system where the ultra-strong coupling regime with the ratio of the coupling strength $g$ to the resonator frequency $\omega$ of order $g/\omega =0.1$ has been achieved~\cite{nd10,fl10,rb12}.
Ultra-strong couplings with a rate up to $g/\omega =0.58$ have been realized with two high-mobility two-dimensional electron gases coupled to a metamaterial~\cite{sm12}. Recently, even higher coupling rates of up to $g/\omega =0.87$ have been reached in semiconductor heterostructures~\cite{ms14}.
The most relevant to the model considered in the present paper are experiments with photonic crystal nanocavities coupled to a short waveguide~\cite{ks13}. Owing to its small optical loss and tight field confinement, waveguides  are capable of mediating strong and long range couplings using photons propagating in their guided modes. Recently, it has been demonstrated experimentally that a strong coupling with a ratio $g/\omega \approx 0.1$ can be achieved between two single-mode cavities subject of a very short decay time of photons out of the cavities to a waveguide composed of discrete modes~\cite{st11}. Schematic diagram of the experiment is shown in~Fig.~\ref{fig1}.

The properties of the coupled modes, including the damping of the modes due to their coupling to the reservoir, are determined by the density operator $\rho$ which satisfies the following master equation
\begin{align}
\frac{d}{dt}\tilde{\rho} = -\frac{i}{\hbar}\left[\tilde{H}_{AB},\tilde{\rho}\right] + {\cal L}\tilde{\rho} ,\label{b1}
\end{align}
where $\tilde{\rho}$ is the density operator in the interaction picture and $\tilde{H}_{AB}$ is the coupling Hamiltonian between the modes
\begin{equation}
\tilde{H}_{AB} = \hbar g\!\left(\hat{a}_{A}^{\dag}\hat{a}_{B}\!+\!\hat{a}_{B}^{\dag}\hat{a}_{A}\!+\!\hat{a}_{A}\hat{a}_{B}e^{2i\omega t}\!+\!\hat{a}_{B}^{\dag}\hat{a}_{A}^{\dag}e^{-2i\omega t}\right) ,\label{b2}
\end{equation}
Taking into account a very short decay time of photons to the waveguide, we have included into the coupling Hamiltonian the resonant (RWA) as well as antiresonant (non-RWA) terms which, as we will see, can have notable contributions at such short evolution times. The RWA terms represent the linear, a beam splitter type coupling between the cavities, whereas the non-RWA terms describes the nonlinear (parametric) type coupling.
In order to distinguish between the contributions of the linear and nonlinear terms, we will work with the Hamiltonian of the form
\begin{align}
\tilde{H}_{AB} &=\hbar\kappa\left(\hat{a}_{A}\hat{a}_{B}^{\dag}+\hat{a}_{A}^{\dag}\hat{a}_{B}\right) \nonumber\\
&+\hbar\epsilon\left(\hat{a}_{A}\hat{a}_{B}e^{2i\omega t} +\hat{a}_{B}^{\dag}\hat{a}_{A}^{\dag}e^{-2i\omega t}\right) ,\label{b4}
\end{align}
where $\kappa$ determines the strength of the linear, whereas $\epsilon$ determines the strength of the nonlinear coupling.
The term ${\cal L}\tilde{\rho}$, appearing in the master equation (\ref{b1}), is an operator representing the damping of the modes to the external environment (reservoir). In general, it contains resonant and antiresonant terms. A recent investigation by Joshi {\it et al.}~\cite{jo14} shows that the antiresonant terms present in the damping part of the master equation can modify the dynamics of strongly coupled modes. However, a further insight into the results reveals that the antiresonant terms change the results quantitatively, but not alter the qualitative behavior. Therefore, we retain only the resonant terms in ${\cal L}\tilde{\rho}$:
\begin{align}
 {\cal L}\tilde{\rho} &= -\frac{1}{2}\sum_{j=A,B}\gamma_{j}\!\left(\hat{a}_{j}^{\dag}\hat{a}_{j}\tilde{\rho} + \tilde{\rho} \hat{a}_{j}^{\dag}\hat{a}_{j} -2\hat{a}_{j}\tilde{\rho} \hat{a}_{j}^{\dag}\right) \nonumber\\
&-\frac{1}{2}\sum_{i\neq j=A,B}\gamma\!\left(\hat{a}_{i}^{\dag}\hat{a}_{j}\tilde{\rho} + \tilde{\rho}\hat{a}_{i}^{\dag}\hat{a}_{j} - 2\hat{a}_{j}\tilde{\rho} \hat{a}_{i}^{\dag}\right) ,\label{b3}
\end{align}
where $\gamma_{j}$ is the damping rate of the mode $j$, and $\gamma$ is the cross damping rate at which the modes are coupled to each other through the interaction with the same reservoir. The coupling reflects the fact that, as  a photon is emitted by the spontaneous decay of the mode $A$ it can be absorbed by the mode $B$, and vice versa. In other words, $\gamma$ describes a collective damping of the modes. The strength of the collective damping depends on the rates $\gamma_{A}$ and $\gamma_{B}$ and the polarization of the modes that $\gamma =\sqrt{\gamma_{A}\gamma_{B}}\cos\theta$, where $\theta$ is the angle between the polarization directions of the modes. If the polarizations are parallel then $\theta =0$ and the collective damping is maximal, $\gamma=\sqrt{\gamma_{A}\gamma_{B}}$, while if the polarizations are perpendicular, then $\gamma=0$.

An obvious question arises, under which conditions both terms in the Hamiltonian (\ref{b4}) could simultaneously contribute to the dynamics of the system. In the presence of the antiresonant terms there are two time scales of the evolution of the system, one determined by the parameters $\kappa, g$ and $\gamma_{j}$ and the other determined by $\omega$. The resonant terms in the master equation (\ref{b1}) experience a variation on a time scale $\Delta t_{r}\sim 1/\kappa, (\sim 1/g, 1/\gamma_{j})$, whereas the antiresonant terms experience a variation on a time scale of $\Delta t_{ar}\sim 1/\omega$.
Therefore, these two time scales should be comparable $(\Delta t_{r}\approx \Delta t_{ar})$ in order the steady state be reached with the antiresonant terms participating fully in the dynamics. Thus, observation (detection) times should be comparable to $\Delta t_{ar}$.

In what follows, we explore the role of the resonant and antiresonant terms on the steady-state characteristics of the system. Analytic expressions are obtained for the density matrix elements which then are used to investigate the influence of the two kind of couplings between the modes on the population distribution, distinguishability and entanglement of the modes.

\section{Steady-state solutions}\label{sec3}

Given the master equation (\ref{b1}), we can use the photon number representation for the density operator and derive equations of motion for the density matrix elements. Suppose that initially there is no excitation present in the modes, i.e., the initial state of the system was a vacuum state $\ket{0_{A}}\ket{0_{B}}$. Since there is no external excitation field present, one would expect that the modes would remain in their vacuum states for all times. However, we will demonstrate that the system evolves to a steady-state in which the singly and doubly excited states can have nonzero populations. To demonstrate this, we consider a basis set of low excitation states consisting of four states
\begin{eqnarray}
\ket{1} &= \ket{0_{A}}\ket{0_{B}} ,\quad \ket{2} =\ket{0_{A}}\ket{1_{B}} , \nonumber\\
\ket{3} &= \ket{1_{A}}\ket{0_{B}} ,\quad \ket{4} =\ket{1_{A}}\ket{1_{B}} ,\label{9}
\end{eqnarray}
where $\ket{0_{j}}$ and $\ket{1_{j}}$ are zero and one excitation states of the cavity $j$. The singly and doubly excited states have been included into the basis in order to fully account effects of the antiresonant terms $\hat{a}_{A}\hat{a}_{B}$ and $\hat{a}_{B}^{\dagger}\hat{a}_{A}^{\dagger}$, which couple the vacuum state to higher excitation states.

The reason for the inclusion of the low excitation states can be understood by noting that the inclusion of the antiresonant terms in the master equation (\ref{b1}) leads to the steady-state to be achieved on a time scale of order $\Delta t \sim 1/\omega$. If the evolution time is of order $\Delta t$, the energy-time uncertainty principle, $\Delta E \Delta t \geq \hbar/2$, enforces that a precision $\Delta E$ of the energy of photons has to be at least of order of $\Delta E \approx \hbar \omega$, which is of order of the one-photon energy. Thus, over the evolution time $\Delta t \approx 1/\omega$, an excitation of the system to the states $\ket{1_{A}}\ket{0_{B}}$, $\ket{0_{A}}\ket{1_{B}}$, and $\ket{1_{A}}\ket{1_{B}}$ is possible.

In the basis (\ref{9}) the density operator $\rho$ has fifteen independent matrix elements. The equations of motion for the density matrix elements which can have nonzero values in the steady state are
\begin{eqnarray}
\dot{\rho}_{11} &=& \gamma_{B}\rho_{22} + \gamma_{A}\rho_{33} + \gamma\left(\rho_{23} + \rho_{32}\right) + i\epsilon\left(\rho_{14} - \rho_{41}\right) ,\nonumber\\
\dot{\rho}_{22} &=&  \gamma_{A} -2\gamma_{0}\rho_{22} - \gamma_{A}\left(\rho_{11} +\rho_{33}\right) \nonumber\\
&& -\frac{1}{2}\left(\gamma - 2i\kappa\right)\rho_{23} -\frac{1}{2}\left(\gamma + 2i\kappa\right)\rho_{32} ,\nonumber\\
\dot{\rho}_{33} &=& \gamma_{B} -2\gamma_{0}\rho_{33} - \gamma_{B}\left(\rho_{11} + \rho_{22}\right) \nonumber\\
&& -\frac{1}{2}\left(\gamma + 2i\kappa\right)\rho_{23} -\frac{1}{2}\left(\gamma - 2i\kappa\right)\rho_{32} ,\nonumber\\
\dot{\rho}_{23} &=& \gamma -\gamma_{0}\rho_{23} - \gamma\left(\rho_{11} +\rho_{22} +\rho_{33}\right) \nonumber\\
&&-\frac{1}{2}\left(\gamma -2i\kappa\right)\rho_{22} -\frac{1}{2}\left(\gamma + 2i\kappa\right)\rho_{33} ,\nonumber\\
\dot{\rho}_{14} &=& -i\epsilon -\left(\gamma_{0} - 2i\omega\right)\rho_{14} +i\epsilon\left(2\rho_{11} + \rho_{22} + \rho_{33}\right) ,\label{9a}
\end{eqnarray}
where $\gamma_{0} = (\gamma_{A}+\gamma_{B})/2$, and $\rho_{44}$ is found from the closure relation of the conservation
of the total population, $\rho_{11} + \rho_{22} + \rho_{33} +\rho_{44} =1$. The set of coupled equations for the density matrix elements involves the populations and the one-photon $\rho_{23}$ and two-photon $\rho_{14}$ coherences.

The set of the differential equations can be written in a matrix form
\begin{align}
\frac{d}{dt}\vec{Y}  = M\vec{Y} + \vec{P} ,\label{v22}
\end{align}
where the vector $\vec{Y}$ has the components
\begin{align}
Y_{1} &= \rho_{11} ,\ Y_{2}=\rho_{22} ,\ Y_{3} =\rho_{33} ,\ Y_{4} = \rho_{23}+\rho_{32} ,\nonumber\\
Y_{5} &= i\!\left(\rho_{23}\!-\!\rho_{32}\right) ,\ Y_{6} = \rho_{14}\!+\!\rho_{41} ,\ Y_{7} = i\!\left( \rho_{14}\!-\!\rho_{41}\right) .
\end{align}
Nonzero components of the vector $\vec{P}$ are
\begin{align}
P_{2} = \gamma_{A}, \ P_{3} =\gamma_{B} ,\ P_{4} = 2\gamma ,\ P_{7} = 2\epsilon ,
\end{align}
and $M$ is the $7\times 7$ matrix of real coefficients
%\begin {widetext}
 \begin{align}
  M = \left(
    \begin{array}{ccccccc}
      0 & \gamma_{B} & \gamma_{A} & \gamma & 0 & 0 & \epsilon \\
      -\gamma_{A} & -2\gamma_{0} & -\gamma_{A} & -\gamma/2 & \kappa & 0 & 0 \\
      -\gamma_{B} & -\gamma_{B} & -2\gamma_{0} & -\gamma/2 & -\kappa & 0 & 0 \\
      -2\gamma & -3\gamma  & -3\gamma & -\gamma_{0} & 0 & 0 & 0 \\
      0 & -2\kappa & 2\kappa & 0 & -\gamma_{0} & 0 & 0 \\
      0 & 0 & 0 & 0 & 0 & -\gamma_{0} & 2\omega \\
      -4\epsilon & -2\epsilon & -2\epsilon & 0 & 0 & -2\omega & -\gamma_{0}
    \end{array}\right) .\label{e19}
\end{align}
%\end{widetext}
The matrix $M$ describes the effects of the coupling terms $\kappa$ and $\epsilon$ as well as those of the dampings.

Solving Eq.~(\ref{v22}) for the steady-state, we find the diagonal matrix elements to be
\begin{align}\label{24}
\rho^{s}_{11} &= \frac{(\epsilon^{2}\!+\!4\omega^2\!+\!\gamma_{0}^2)}{D}\!\left[4\kappa^{2}\!\left(\gamma_{0}^{2} -\gamma^{2}\right)\!+\!\gamma_{0}^{2}\!\left(\gamma_{A}\gamma_{B} -\gamma^{2}\right)\right] ,\nonumber \\
\rho^{s}_{22} &= \frac{\epsilon^2}{D}\left[\left(4\kappa^2+\gamma_A^2\right)\left(\gamma_{0}^2- \gamma^2\right) +\frac{1}{4}\gamma^2\left(\gamma_A-\gamma_B\right)^2\right] , \nonumber\\
\rho^{s}_{33} &= \frac{\epsilon^2}{D}\left[(4\kappa^2+\gamma_B^2)\big(\gamma_{0}^2-\gamma^2\big)+\frac{1}{4}\gamma^2(\gamma_A-\gamma_B)^2\right] , \nonumber\\
\rho^{s}_{44} &= \frac{\epsilon^2}{D}\left[4\kappa^{2}\left(\gamma_{0}^{2} -\gamma^{2}\right) +\gamma_{0}^{2}\left(\gamma_{A}\gamma_{B} -\gamma^{2}\right)\right] ,
\end{align}
and the off-diagonal elements
\begin{align}\label{24u}
\rho^{s}_{14} &= \frac{i\epsilon(\gamma_{0}+2i\omega)}{D}\left[4\kappa^{2}\left(\gamma_{0}^{2} -\gamma^{2}\right) +\gamma_{0}^{2}\left(\gamma_{A}\gamma_{B} -\gamma^{2}\right)\right] ,\nonumber \\
\rho^{s}_{23} &= \frac{i(\gamma_{A}\!-\!\gamma_B)\epsilon^2}{4D}\!\left[8\kappa\big(\gamma_{0}^2-\gamma^2\big) +i\gamma(\gamma_A^2-\gamma_B^2)\right] ,
\end{align}
where
\begin{align}\label{25}
 D &= \left(\gamma_{0}^{2}+4\omega^{2}\right)\left[4\kappa^{2}\left(\gamma_{0}^{2}-\gamma^{2}\right) +\gamma_{0}^{2}\left(\gamma_{A}\gamma_{B}-\gamma^{2}\right)\right] \nonumber\\
 &+4\epsilon^{2}\left(\gamma_{0}^{2}-\gamma^{2}\right)\left(\gamma_{0}^{2} +4\kappa^{2}\right) .
\end{align}
From Eq.~(\ref{24}), we see that the steady-state of the coupled modes is not the ground state $\ket 1 = \ket{0_{A}}\ket{0_{B}}$. The population is redistributed between the states including the doubly excited state $\ket 4$. There are no external sources of photons, like driving laser fields in the system. This effect occurs passively by just adding the antiresonant (non-RWA) terms determined by $\epsilon$. When these terms are ignored, the standard RWA result is obtained with $\rho_{11}=1$ and no population in the excited states.

In addition, the steady-state solution is strongly affected by the coupling of the modes to the reservoir. In particular, a coherence is generated in the process of spontaneous emission with unequal damping rates, $\gamma_{A}\neq \gamma_{B}$. Let us discuss in greater detail the cases of independent $(\gamma =0)$ and collective $(\gamma\neq 0)$ dampings, under unbalanced $(\gamma_{A}\neq \gamma_{B})$ and balanced $(\gamma_{A}=\gamma_{B})$ decays of the modes.

\subsection{Unbalanced decay: $\gamma_{A}\neq \gamma_{B}$}

 When the modes decay independently, $\gamma =0$, and then the steady-state solution~(\ref{24}) reduces to
 \begin{eqnarray}
\rho^{s}_{11} &=& \frac{\left(4\kappa^2+\gamma_A\gamma_B\right)}{D_{0}}\left(4\omega^2+\epsilon^2+\gamma_{0}^2\right) ,\nonumber\\
\rho^{s}_{22} &=& \frac{\epsilon^2\left(4\kappa^2+\gamma_A^2\right)}{D_{0}}, \quad
\rho^{s}_{33} = \frac{\epsilon^2\left(4\kappa^2+\gamma_B^2\right)}{D_{0}}, \nonumber\\
\rho^{s}_{44} &=& \frac{\epsilon^2\left(4\kappa^2+\gamma_A\gamma_B\right)}{D_{0}} ,\quad
\rho^{s}_{23} = \frac{2i(\gamma_A-\gamma_B)\kappa\epsilon^2}{D_{0}} ,\nonumber\\
\rho^{s}_{14} &=& \frac{i\epsilon\left(4\kappa^2+\gamma_A\gamma_B\right)}{D_{0}}(\gamma_{0}+2i\omega) ,\label{18}
\end{eqnarray}
where
\begin{eqnarray}
 D_{0} = \left(4\kappa^2\!+\!\gamma_A\gamma_B\right)\!\left(4\omega^2\!+\!\gamma_{0}^2\right)
 + 4\epsilon^{2}\!\left(4\kappa^2\!+\!\gamma_{0}^2\right) .\label{19}
\end{eqnarray}
Expression for $\rho_{23}^{s}$ shows that a coherence is generated by spontaneous decay of the modes even if the modes decay independently. It requires the modes to decay with unequal rates, $\gamma_{A}\neq \gamma_{B}$; that is, unbalanced decay plays a constructive role in the generation of the one-photon coherence. The unbalanced decay of the modes creates a population inversion between states $\ket 2$ and $\ket 3$ that
\begin{equation}
\rho^{s}_{22} - \rho^{s}_{33} = \frac{\epsilon^2\left(\gamma_A^2 -\gamma_{B}^{2}\right)}{D_{0}} = \frac{2\epsilon^{2}\gamma_{0}\left(\gamma_A -\gamma_{B}\right)}{D_{0}} .
\end{equation}
Then the coherence can be written as
\begin{equation}
\rho^{s}_{23} = \frac{2i(\gamma_A-\gamma_B)\kappa\epsilon^2}{D_{0}} = \frac{i\kappa}{\gamma_{0}}\left(\rho^{s}_{22} - \rho^{s}_{33}\right) .
\end{equation}
This shows the familiar fact that the coherence between two states is proportional to the product of the driving field strength and the population inversion. That is, the linear coupling $\kappa$ between the modes is a complete analog of a coherent driving of two quantum states.

Another interesting observation is that the unbalanced decay can lead to a population inversion between the doubly excited state $\ket 4$ and the singly excited states $\ket{2}$ and $\ket{3}$). Really, if we evaluate ratios $\rho^{s}_{22}/\rho^{s}_{44}$ and $\rho^{s}_{33}/\rho^{s}_{44}$, we find the result
\begin{align}
\frac{\rho^{s}_{22}}{\rho^{s}_{44}} &= 1 + \frac{\gamma_{A}\left(\gamma_{A}-\gamma_{B}\right)}{\kappa^{2} +\gamma_{A}\gamma_{B}} ,\nonumber\\
\frac{\rho^{s}_{33}}{\rho^{s}_{44}} &= 1- \frac{\gamma_{B}\left(\gamma_{A}-\gamma_{B}\right)}{\kappa^{2} +\gamma_{A}\gamma_{B}} .
\end{align}
We see that depending on whether $\gamma_{A}>\gamma_{B}$ or  $\gamma_{A}<\gamma_{B}$, the population can be inverted between $\ket 4$ and either $\ket 2$ or $\ket 3$. It is interesting that the population can be inverted between $\ket 4$ and only one of the singly excited states.

Consider now the case when the modes decay collectively. If the collective damping rate is maximal, $\gamma =\sqrt{\gamma_{A}\gamma_{B}}$, the solution (\ref{24}) simplifies to
\begin{align}\label{24a}
\rho^{s}_{11} &= \frac{\kappa^{2}\left(\epsilon^{2} + 4\omega^2 + \gamma_{0}^2\right)}{ \tilde{D}} ,\nonumber \\
\rho^{s}_{22} &= \frac{\epsilon^{2}\left(2\kappa^2+\gamma_{A}\gamma_{0}\right)}{2 \tilde{D}} ,\quad
\rho^{s}_{33} = \frac{\epsilon^{2}\left(2\kappa^2+\gamma_{B}\gamma_{0}\right)}{2 \tilde{D}} , \nonumber\\
\rho^{s}_{44} &= \frac{\kappa^{2}\epsilon^2}{ \tilde{D}} ,\quad
\rho^{s}_{14} = \frac{i\epsilon\kappa^{2}\left(\gamma_{0}+2i\omega\right)}{\tilde{D}} ,\nonumber \\
\rho^{s}_{23} &= \frac{i\epsilon^2}{2\tilde{D}} \left[\kappa\left(\gamma_{A} -\gamma_{B}\right) + i\gamma_{0}\sqrt{\gamma_{A}\gamma_{B}}\,\right] ,
\end{align}
with
\begin{align}\label{25a}
 \tilde{D} &= \kappa^{2}\left(\gamma_{0}^{2}+4\omega^{2}\right) + \epsilon^{2}\left(\gamma_{0}^{2}+4\kappa^{2}\right) .
\end{align}
There are several important differences between Eq.~(\ref{24a}) and the result (\ref{18}) for independent reservoirs.

First of all, the coherence $\rho_{23}^{s}$ is composed of two parts: the part proportional to $\kappa$ is driven directly by the linear coupling between the modes, while the part proportional to $\gamma$ results from an exchange of the excitation through the coupling of the modes to the same reservoir. This shows that a coherence between two states can be generated even if there is no population difference between the states. This property of the coherence can have an interesting effect on the redistribution of the population between the states. It is easily seen from Eq.~(\ref{24a}) that in the absence of the linear coupling $(\kappa =0)$ the entire population is redistributed (trapped) in the single excitation states with the populations of the states and the coherence between them given by
\begin{equation}
\rho^{s}_{22} = \frac{\gamma_{A}}{2\gamma_{0}} ,\quad
\rho^{s}_{33} = \frac{\gamma_{B}}{2\gamma_{0}} , \quad
\rho^{s}_{23} = -\frac{\sqrt{\gamma_{A}\gamma_{B}}}{2\gamma_{0}} .\label{b25}
\end{equation}
We may introduce symmetric and antisymmetric combinations of the singly excitation states
\begin{eqnarray}
\ket{b} &=& \frac{1}{\sqrt{2\gamma_{0}}}\left(\sqrt{\gamma_{A}}\ket 3 + \sqrt{\gamma_{B}}\ket 2 \right) ,\nonumber\\
\ket{d} &=& \frac{1}{\sqrt{2\gamma_{0}}}\left(\sqrt{\gamma_{B}}\ket 3 - \sqrt{\gamma_{A}}\ket 2 \right) ,\label{b22}
\end{eqnarray}
and find using Eq.~(\ref{b25}) that $\rho_{bb}=0$ and $\rho_{dd}=1$. Clearly, the steady-state of the modes is not a mixed state but a pure entangled state $\ket d$. Thus, despite the interaction with a dissipative reservoir, the system evolves to a pure entangled state rather than the expected mixed state.

In addition, there is no population inversion between the double excitation state $\ket 4$ and the single excitation states $\ket 2$ and $\ket 3$. It is easy to see, Eq.~(\ref{24a}) for the populations lead to ratios
\begin{equation}
\frac{\rho^{s}_{22}}{\rho^{s}_{44}} = 1 + \frac{\gamma_{A}\gamma_{0}}{2\kappa^{2}} ,\quad
\frac{\rho^{s}_{33}}{\rho^{s}_{44}} = 1+ \frac{\gamma_{B}\gamma_{0}}{2\kappa^{2} } ,
\end{equation}
which are always greater than $1$.

\subsection{Balanced decay: $\gamma_{A} =\gamma_{B}$}\label{sec3b}

Let us now discuss the steady-state solutions in the case of balanced decay of the modes, i.e., decay with equal damping rates, $\gamma_{A}=\gamma_{B}$. We will see that this leads to quite different features than those found for unbalanced decays. The most important difference is that it requires to consider separately the steady-state solutions for two regions of $\gamma$: $\gamma < \gamma_{0}$ and $\gamma =\gamma_{0}$. This is because the determinant of the matrix $M$, Eq.~(\ref{e19}), is equal to zero when  $\gamma=\sqrt{\gamma_{A}\gamma_{B}}$ and $\gamma_{A}=\gamma_{B}$.

We first examine the steady-state solution for $\gamma <\gamma_{0}$.
\begin{align}\label{24p}
\rho^{s}_{11} &= \frac{\epsilon^{2} + \gamma_{0}^{2} + 4\omega^2}{D^{\prime}} ,\quad
\rho^{s}_{22} = \rho^{s}_{33} = \rho^{s}_{44} = \frac{\epsilon^2}{D^{\prime}} ,\nonumber\\
\rho^{s}_{14} &= \frac{i\epsilon(\gamma_{0}+2i\omega)}{D^{\prime}} ,\quad
\rho^{s}_{23} = 0 ,
\end{align}
where $D^{\prime} = 4\epsilon^{2} +\gamma_{0}^{2}+4\omega^{2}$. We see that as long as $\gamma <\gamma_{0}$, the system relaxes to a mixed state which is independent of $\gamma$ and $\kappa$. Moreover, the populations of the singly and doubly excited states are exactly equal. In other words, when measuring the populations of the excited states, all measurement outcomes would occur with equal probability. Since $\rho^{s}_{23}=0$, no entangled states are created between the singly excited states. We can conclude that as long as $\gamma_{A}=\gamma_{B}\equiv \gamma_{0}$ and $\gamma <\gamma_{0}$, there is no difference in the decay of the modes into local reservoirs and into a common reservoir.

The fact that the result (\ref{24p}) is independent of $\gamma$ may lead one to conclude that it is also valid in the limit of $\gamma =\gamma_{0}$.
But this result is {\it not} correct in this limit since Det$[M]=0$ when $\gamma=\sqrt{\gamma_{A}\gamma_{B}}$ and $\gamma_{A}=\gamma_{B}\equiv \gamma_{0}$. In order to find the correct steady-state of the system, we rewrite the equations of motion (\ref{9a}) in the basis, $\{\ket{1}, \ket{b}, \ket{d}, \ket{4}\}$ and find that the corresponding equations of motion are
\begin{eqnarray}
\dot{\rho}_{dd} &=&  0 ,\nonumber\\
\dot{\rho}_{bb} &=&  2\gamma_{0}\!\left(1\!-\!\rho_{dd}\right) -2\gamma_{0}\rho_{bb} -2\gamma_{0}\rho_{11} ,\nonumber\\
\dot{\rho}_{bd} &=&  -\left(\gamma_{0} + 2i\kappa\right)\rho_{bd} ,\nonumber\\
\dot{\rho}_{11} &=& 2\gamma_{0}\rho_{bb} + i\epsilon\left(\rho_{14} - \rho_{41}\right) ,\nonumber\\
\dot{\rho}_{14} &=& -i\epsilon -\left(\gamma_{0}\!-\!2i\omega\right)\!\rho_{14} + i\epsilon\!\left(2\rho_{11}\!+\!\rho_{bb}\!+\!\rho_{dd}\right) .\label{25e}
\end{eqnarray}
Since $\dot{\rho}_{dd} = 0$, the state $\ket d$ is totally decoupled from the remaining states and does not evolve in time. In other words, an initial population of the state $\ket d$ will remain constant for all times.

With $\rho_{dd}$ constant, the steady-state solution of Eq.~(\ref{25e}) is of the form
\begin{eqnarray}
\rho_{bd} &=& \rho_{db} = 0 ,\nonumber\\
\rho_{dd} &=& \rho_{dd}(0) ,\nonumber\\
\rho_{bb} &=& \frac{\epsilon^2}{3\epsilon^2+\gamma_{0}^2+4\omega^2}\left[1-\rho_{dd}(0)\right] ,\nonumber\\
\rho_{11} &=& \frac{\epsilon^2+\gamma_{0}^2+4\omega^2}{3\epsilon^2+\gamma_{0}^2+4\omega^2}\left[1-\rho_{dd}(0)\right] ,\nonumber\\
\rho_{14} &=& i\frac{\epsilon(\gamma_{0} +2i\omega)}{3\epsilon^2+\gamma_{0}^2+4\omega^2}\left[1-\rho_{dd}(0)\right] .\label{26a}
\end{eqnarray}
In terms of the product state basis, the corresponding solution is
\begin{eqnarray}
  \rho_{11}^{s} &=& \frac{\epsilon^2+\gamma_{0}^2+4\omega^2}{3\epsilon^2+\gamma_{0}^2+4\omega^2}\left[1-\rho_{dd}(0)\right] ,\nonumber\\
  \rho_{22}^{s} &=& \rho_{33}^{s} = \frac{1}{2}\left\{1 -\frac{2\epsilon^2+\gamma_{0}^2+4\omega^2}{3\epsilon^2+\gamma_{0}^2+4\omega^2}
  \left[1-\rho_{dd}(0)\right]\right\} ,\nonumber\\
  \rho_{44}^{s} &=& \frac{\epsilon^2}{3\epsilon^2+\gamma_{0}^2+4\omega^2}\left[1-\rho_{dd}(0)\right] ,\nonumber\\
   \rho_{14}^{s}  &=& i\frac{\epsilon(\gamma_{0}+2i\omega)}{3\epsilon^2+\gamma_{0}^2+4\omega^2}\left[1-\rho_{dd}(0)\right] ,\nonumber\\
 \rho_{23}^{s}  &=& -\frac{1}{2}\left\{1 -\frac{4\epsilon^2+\gamma_{0}^2+4\omega^2}{3\epsilon^2+\gamma_{0}^2+4\omega^2}\left[1-\rho_{dd}(0)\right]\right\} .\label{26b}
\end{eqnarray}
We see that the physical consequences of the complete decoupling of the state $\ket d$ from the remaining states is the dependence of the steady-state values of the density matrix elements on initial conditions. Note that the system no longer evolves to a pure state $\it unless$ it is prepared initially in the state $\ket d$.
Thus, depending on the way we prepare the system initially, we can realize different situations. Note also the steady-state of the system is independent of~$\kappa$. Hence, if $\rho_{dd}(0)=1$, the only steady-state for Eq.~(\ref{25e}) is $\rho_{dd}=1$ with all other density matrix elements equal to zero. It means that if the system is initially prepared in the state $\ket d$, it will remain in this state for all times, i.e., $\rho_{dd}(t) = \rho_{dd}(0)$.

\section{Distinguishability of the modes}\label{sec4}

The presence of the linear and nonlinear couplings between the cavities $A$ and $B$ may lead one to suspect that the modes of the cavities are indistinguishable. In particular, if we assume that only a single excitation is present that the system is in either $\ket{1_{A}}\ket{0_{B}}$ or $\ket{0_{A}}\ket{1_{B}}$ state, then the action of the linear coupling $\kappa$ generates a state which is a linear superposition of the one-photon states. As is well known, the probability of detecting a photon emitted from the superposition state exhibits interference effects. The interference is regarded as a signature of indistinguishability of the states.

Nevertheless, we will demonstrate that the modes can be distinguishable even in the presence of the couplings that "which-path" information is made possible due to the inclusion of the state $\ket{1_{A}}\ket{1_{B}}$ enforced by the energy-time uncertainty principle. However, the "which-way" information can be erased by allowing the cavities to decay with different rates. To show this, we consider electromagnetic fields $\hat{E}_{A}(\vec{r},t)$ and $\hat{E}_{B}(\vec{r},t)$ of the cavities $A$ and $B$ at position $\vec{r}$ at time $t$. Since fields of the cavities are treated as single-mode fields, the negative frequency parts of the fields can be written as
\begin{align}
\hat{E}^{(-)}_{A}(\vec{r},t) &= {\cal E}\hat{a}_{A}e^{i\left(\vec{k}_{A}\cdot \vec{r} - \omega t\right)} ,\nonumber\\
\hat{E}^{(-)}_{B}(\vec{r},t) &= {\cal E}\hat{a}_{B}e^{i\left(\vec{k}_{B}\cdot \vec{r} - \omega t\right)} ,
\end{align}
where $\vec{k}_{A}$ and $\vec{k}_{B}$ are wave vectors of the modes and ${\cal E}$ is a constant amplitude.
Then the intensity of the field detected by a photodetector located at $\vec{r}$ at time $t$ is given~by
\begin{align}
I(\vec{r},t) &= \alpha \left\langle \left(\hat{E}^{(+)}_{A} + \hat{E}^{(+)}_{B}\right)\left(\hat{E}^{(-)}_{A} + \hat{E}^{(-)}_{B}\right)\right\rangle \nonumber\\
&= \alpha|{\cal E}|^{2}\left\{2\rho_{44} + \rho_{22} + \rho_{33}\right. \nonumber\\
&\left. +\, 2|\rho_{23}|\cos\!\left[\left(\vec{k}_{A}\!-\!\vec{k}_{B}\!\right)\!\cdot\!\vec{r} +{\rm arg}(\phi_{A}\!-\!\phi_{B})\right]\right\} ,\label{b37}
\end{align}
where $\alpha$ is a constant characteristic of the detector, and we have written
\begin{align}
\langle \hat{a}^{\dag}_{A}\hat{a}_{A}\rangle &= \rho_{44} + \rho_{22} ,\quad \langle \hat{a}^{\dag}_{B}\hat{a}_{B}\rangle = \rho_{44} + \rho_{33} ,\nonumber\\
\langle \hat{a}^{\dag}_{A}\hat{a}_{B}\rangle &=  \langle \hat{a}^{\dag}_{B}\hat{a}_{A}\rangle^{\ast} = |\rho_{23}|e^{i(\phi_{A}-\phi_{B})} .
\end{align}
We see from Eq.~(\ref{b37}) that the intensity varies periodically with position only if the coherence $|\rho_{23}|$ is different from zero. From the definition of the first-order visibility and Eq.~(\ref{b37}), we find that
\begin{align}
{\cal V} = \frac{I_{max} - I_{min}}{I_{max}+I_{min}} = \frac{2|\rho_{23}|}{2\rho_{44} + \rho_{22}+\rho_{33}} ,\label{b39a}
\end{align}
and then by using Eqs.~(\ref{24}) and (\ref{24u}) we find
\begin{align}
{\cal V} = \frac{|\gamma_{d}|\sqrt{4\kappa^{2}(\gamma_{0}^{2}-\gamma^{2})^{2} +(\gamma\gamma_{0}\gamma_{d})^{2}}}{(4\kappa^{2}+\gamma_{0}^{2})(\gamma_{0}^{2}-\gamma^{2})} ,\label{b39}
\end{align}
where $\gamma_{d}=(\gamma_{A}-\gamma_{B})/2$.
This simple result for the first-order visibility is strongly dependent on whether the modes are damped with equal $(\gamma_{A}=\gamma_{B})$ or unequal $(\gamma_{A}\neq \gamma_{B})$ rates. If the modes are damped with equal rates, $\gamma_{d}=0$, and then the interference pattern vanishes. Hence, independent of the presence of the couplings, the modes are distinguishable when are damped with the same rates. The reason of the distinguishability of the modes is the inclusion of the state $\ket{1_{A}}\ket{1_{B}}$ into the dynamics of the system enforced by the energy-time uncertainty principle. In physical terms, we may attribute this to the fact that the modes, each occupied by a photon, are resolved at the detector. For example, if a photon is detected in mode $A$ it must come from this mode since two occupied modes cannot exchange photons. An alternative explanation is that two decay channels from the state $\ket{1_{A}}\ket{1_{B}}$ exist: $\ket{1_{A}}\ket{1_{B}}\rightarrow \ket{0_{A}}\ket{1_{B}}$ and $\ket{1_{A}}\ket{1_{B}}\rightarrow \ket{1_{A}}\ket{0_{B}}$. Then, one can distinguish from which channel the detected photon came by measuring the population of the states $\ket{1_{A}}\ket{0_{B}}$ and $\ket{0_{A}}\ket{1_{B}}$.

One can notice from Eq.~(\ref{b39}) that the visibility is independent of $\epsilon$. Note also that for the visibility to be nonzero it is required that not only $\gamma_{d}\neq 0$ but also $\kappa\neq 0$ and/or $\gamma\neq 0$. Thus, in the case of the collective decay $(\gamma\neq 0)$ the visibility can be different from zero even if $\kappa =0$.
Equation~(\ref{b39}) also shows that the visibility is maximal when either $\gamma_{A}\gg\gamma_{B}$ or $\gamma_{B}\gg\gamma_{A}$.
Consequently, we can make the modes indistinguishable by erasing one of the photons through a fast spontaneous emission of one of the two modes.
To put it another way, when one of the photons is erased by spontaneous emission then the remaining photon can produce the interference since in the presence of the coupling $\kappa$ it is impossible to determine from which mode the detected photon came. This restores the first-order interference which is a manifestation of the intrinsic indistinguishability of two possible paths of the detected photon.

It is worth emphasizing that there is an upper limit of $50\%$ for the first-order visibility ${\cal V}$ when the modes independently decay to the reservoirs. On the other hand, when the modes decay collectively the visibility can be close to unity and can be independent of $\kappa$. To show this, we introduce ratios $R\equiv \kappa/\gamma_{0}$ and $u\equiv \gamma_{d}/\gamma_{0}$, and then find that Eq.~(\ref{b39}) yields
\begin{align}
{\cal V}\equiv {\cal V}_{s} = |u|\frac{2R}{4R^{2}+1} \label{b41}
\end{align}
for the decay to separate reservoirs $(\gamma=0)$, and
\begin{align}
{\cal V}\equiv {\cal V}_{c} = \frac{\sqrt{4R^{2}u^{2} + 1 -u^{2}}}{4R^{2}+1} \label{b42}
\end{align}
for the decay to a common reservoir with $\gamma = \sqrt{\gamma_{A}\gamma_{B}}$.

Since $|u|\leq 1$, the visibility ${\cal V}_{s}$ can be no larger than $50\%$, and it is required $R\neq 0$ for ${\cal V}_{s}$ to be different from zero. It follows from Eq.~(\ref{b41}) that ${\cal V}_{s}$ has its largest value of ${\cal V}_{s}=1/2$ when $R=1/2$ and $|u|=1$. Clearly, there is an upper limit of $50\%$ for the visibility when the modes decay to separate reservoirs. In contrast, the visibility ${\cal V}_{c}$ can exceed $50\%$ and can approach $100\%$ even when $R=0\, (\kappa =0)$. It can happen when the linear coupling $\kappa$ is weak, $R\ll 1$, or even if it is absent, $R=0$. In this limit, ${\cal V}\approx\sqrt{1-u^{2}}$, which can be close to $1$ when $u\approx 0\, (\gamma_{A}\approx \gamma_{B})$. It should be noted that in this case the system is in the pure entangled state $\ket d$, Eq.~(\ref{b22}), which is not the maximally entangled state. We stress that it is impossible to put $u=1$ in Eq.~(\ref{b42}), at which the visibility would correspond to that of a maximally entangled state since we cannot assume $\gamma_{A}=\gamma_{B}$ in the expression (\ref{b42}). The expression for the visibility given in Eq.~(\ref{b39}) is valid only for $\gamma_{A}\neq \gamma_{B}$.

To consider the limit $\gamma=\sqrt{\gamma_{A}\gamma_{B}}$ with $\gamma_{A}=\gamma_{B}$ in the evaluation of the visibility ${\cal V}$ given by Eq.~(\ref{b39a}), we must apply the steady-state solutions given in Eq.~(\ref{26b}). Thus, substituting Eq.~(\ref{26b}) into Eq.~(\ref{b39a}) we get
\begin{align}
{\cal V} = \frac{|\epsilon^{2}-(4\epsilon^{2}+\gamma_{0}^{2}+4\omega^{2})\rho_{dd}(0)|}{3\epsilon^{2}+(\gamma_{0}^{2}+4\omega^{2})\rho_{dd}(0)} .\label{b40}
\end{align}
In comparison with Eq.~(\ref{b39}), we see that the dependence of the visibility on $\kappa$ is absent. The most obvious difference is the dependence on the initial state $\rho_{dd}(0)$. For $\rho_{dd}(0)=0$, the visibility ${\cal V}=1/3$ irrespective of $\epsilon$ and $\gamma_{0}$. In the other extreme when $\rho_{dd}(0)=1$, the visibility reaches its maximal value of ${\cal V}=1$ also irrespective of $\epsilon$ and $\gamma_{0}$. This behavior can be explained as a result of the transition of the system from a mixed state involving three states $\ket 1, \ket b, \ket 4$ to a pure state involving the state $\ket d$, which is a maximally entangled state. These results suggest that the interference can be used to detect one-photon entangled states in the system.

Finally, we would like to comment about the connection between indistinguishability and the presence of two significantly different decay rates in the system. Although the modes $A$ and $B$ decay with the same rate it must not be thought that in this case the interference pattern is always absent. If the modes decay collectively there are two superposition states in the system $\ket b$ and $\ket d$ which decay with significantly different rates. According to Eq.~(\ref{25e}), the state $\ket b$ decays with a rate $2\gamma_{0}$ whereas the state $\ket d$ is metastable. Clearly, the decay rates of the superposition states are significantly different even when $\gamma_{A}=\gamma_{B}$.
Therefore, we may conclude that the one-photon interference results from the presence of unequal decay rates in the system.

\section{Entanglement between the modes}\label{sec5}

The strong dependence of the steady-state of the system on whether $\gamma_{A}\neq \gamma_{B}$ or $\gamma_{A}=\gamma_{B}$ may have a significant effect on entanglement between the modes. The question of the creation of entanglement between the modes is addressed by considering the concurrence, a measure of entanglement between two systems~\cite{W98}. Since the evolution of the system is described by the density operator whose the matrix representation in the basis (\ref{9}) is of the $X$ form, the concurrence can be calculated analytically and can be expressed as
\begin{equation}\label{10}
    \mathcal{C}(t) = \max\left\{0,C_{1}(t),C_{2}(t)\right\} ,
\end{equation}
where
\begin{eqnarray}
  C_{1}(t) &=& 2\left[|\rho_{23}(t)|-\sqrt{\rho_{11}(t)\rho_{44}(t)}\right]\label{11} ,\label{11}\\
   C_{2}(t) &=& 2\left[|\rho_{14}(t)|-\sqrt{\rho_{22}(t)\rho_{33}(t)}\right] .\label{12}
\end{eqnarray}
There are two quantities which determine a nonzero concurrence. Obviously, either $C_{1}(t)>0$ or $C_{2}(t)>0$ is required for the modes to be entangled. The quantity $C_{1}(t)$ determines an entanglement created by the coherence $\rho_{23}(t)$, whereas $C_{2}(t)$ determines an entanglement created by the coherence $\rho_{14}(t)$. It follows that $C_{1}(t)>0$ corresponds to an entangled state involving the one-photon states while $C_{2}(t)>0$ corresponds to an entangled state involving the zero and two-photon states. We have already seen that the coherence $\rho_{23}(t)$ can be created by the linear coupling $\kappa$ and also by the collective damping $\gamma$, while the coherence $\rho_{14}(t)$ can be created by the nonlinear coupling~$\epsilon$.

\begin{figure}[h]
\center{\includegraphics[width=\columnwidth]{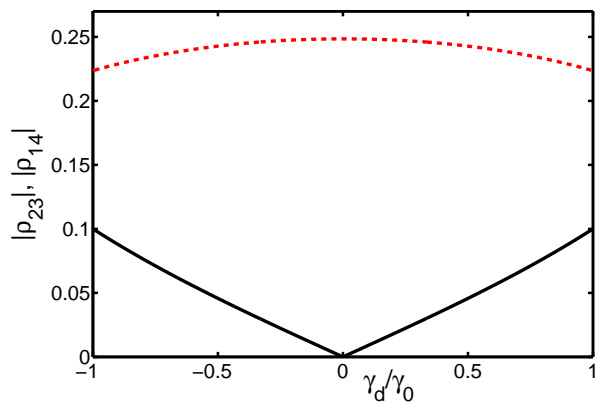}}
\caption{(Color online) Variation of the steady-state coherences $|\rho_{23}|$ (solid black line) and $|\rho_{14}|$ (dashed red line) with $\gamma_{d}$ when the modes decay to separate reservoirs $(\gamma =0)$. The other parameters are $\kappa/\omega =\epsilon/\omega =\gamma_{0}/\omega =1$.}
\label{fig2}
\end{figure}

Although there is no direct connection between the one- and two-photon coherences, we find that in the system considered here the modes exhibit an interesting coherence effect~\cite{ow90,mg96,lm98}. Namely, the modes can be {\it anticoherent} that the one-photon coherence $\rho_{23}$ vanishes and at the same time $\rho_{14}$ is maximal. This is shown in Fig.~\ref{fig2} where we plot the variation of the absolute values of the coherences $|\rho_{23}|$ and $|\rho_{14}|$ with $\gamma_{d}$. When $\gamma_{d}=0$, the coherence $|\rho_{23}|$ vanishes whereas $|\rho_{14}|$ has a maximum. Thus, for $\gamma_{d}=0$ the modes are completely anticoherent. However, as soon as  $\gamma_{d}\neq 0$, the coherence $\rho_{23}$ is different from zero. In this case, the modes are regarded as partially mutually coherent. It is interesting to note from Fig.~\ref{fig2} that an increase of $|\rho_{23}|$ results in a decrease of $|\rho_{14}|$ and vice versa. This "anticoherence" can be reflected in entanglement that an increase of $C_{1}(t)$ leads to a decrease of $C_{2}(t)$.

\subsection{The case of independent decay, $\gamma =0$}

Let us turn to detailed analysis of the concurrence for the case of independent unbalanced decays, $\gamma =0$ and $\gamma_{A}\neq \gamma_{B}$.
Using the steady-state solution (\ref{18}), the concurrence can be easily determined and is given by
\begin{equation}\label{20}
    \mathcal{C}^{s} = \max\left(0,C_{1}^{s},C_{2}^{s}\right) ,
\end{equation}
where
\begin{align}\label{21}
C_{1}^{s} = \frac{2\epsilon}{D_{0}}\!\left[4|\gamma_{d}|\kappa\epsilon -(4\kappa^{2}\!+\!\gamma_{0}^{2}-\gamma_{d}^{2})\sqrt{4\omega^2 +\epsilon^{2}+\gamma_{0}^2}\right] ,
\end{align}
and
\begin{align}\label{22}
 C_{2}^{s} &= \frac{2\epsilon}{D_{0}}\Bigg[(4\kappa^2+\gamma_A\gamma_B)\sqrt{4\omega^2+\gamma_{0}^2} \nonumber\\
&-\epsilon\sqrt{(4\kappa^2+\gamma_A^2)(4\kappa^2+\gamma_B^2)}\Bigg] .
\end{align}
It is seen from Eqs.~(\ref{21}) and (\ref{22}) that a nonzero $\epsilon$ is necessary for both quantities $C^{s}_{1}$ and $C^{s}_{2}$ to be nonzero. However, a nonzero $\kappa$ is needed for $C^{s}_{1}$ to be positive, while $C^{s}_{2}$ can be positive even for $\kappa =0$. Moreover, the damping rates should be different $(\gamma_{A}\neq \gamma_{B})$ for $C^{s}_{1}$ to be positive. This means that in the case of an unbalanced damping rates, entanglement between the modes can be determined by two criteria. These two criteria do not overlap that they determine two separate ranges of the parameters at which entanglement occurs.
\begin{figure}[h]
%\begin{flushleft}\quad\quad\quad\textbf{(a)}\end{flushleft}
\hspace*{-6.3cm}\textbf{(a) }
\\ \textbf{(b)}
\includegraphics[width=.72\columnwidth]{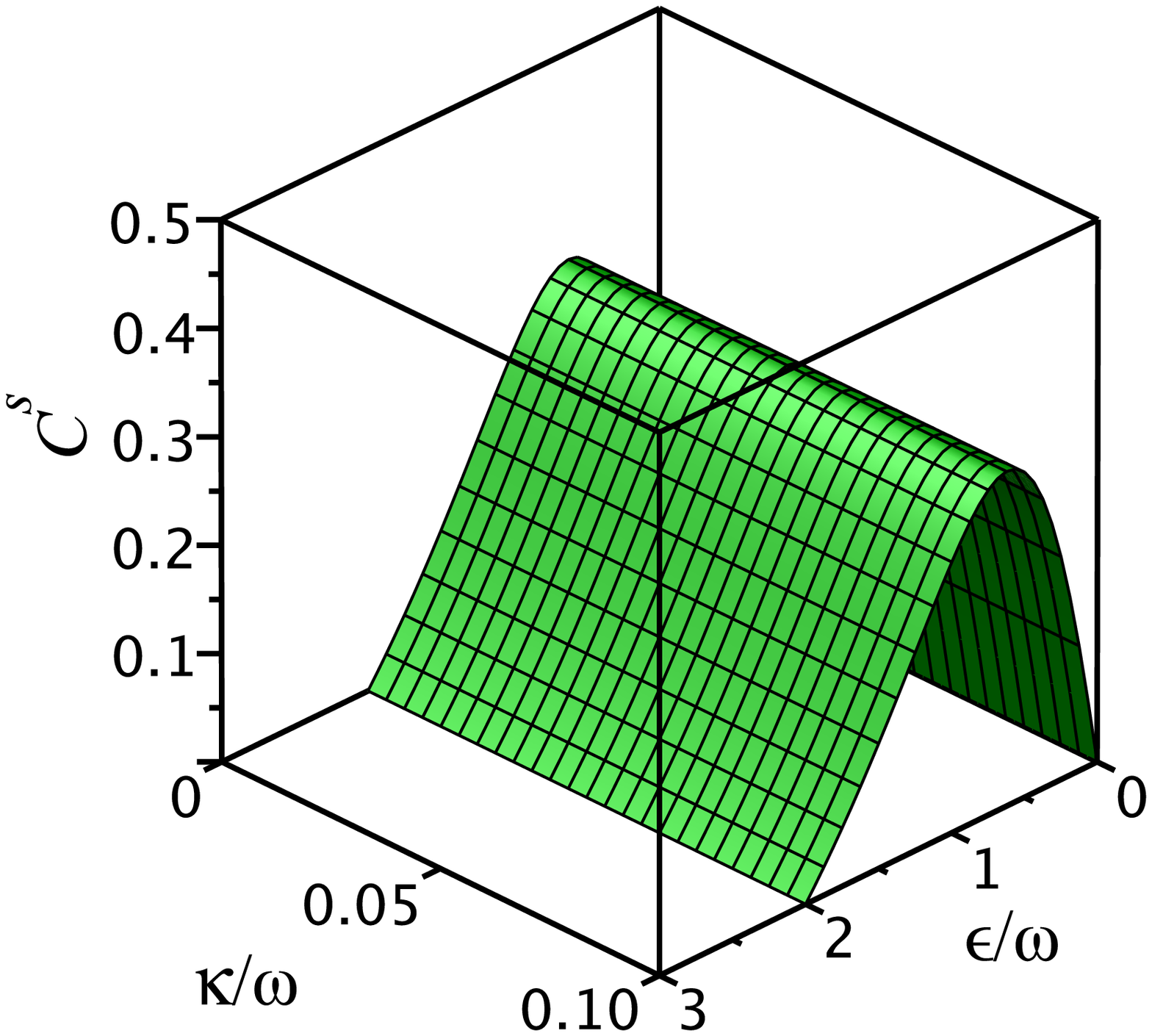}
\\\textbf{(c)}
\includegraphics[width=.72\columnwidth]{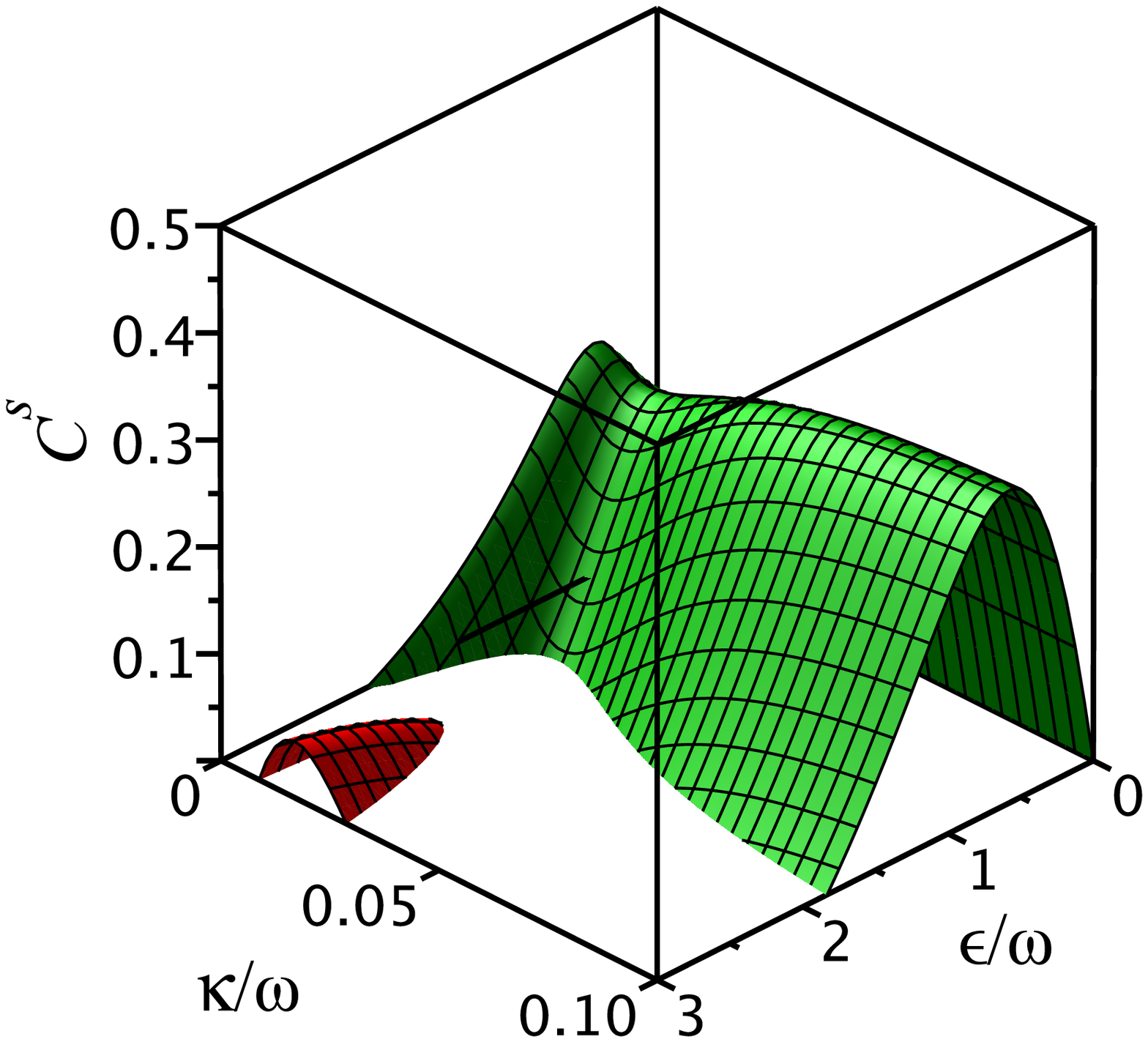}\\
\includegraphics[width=.72\columnwidth]{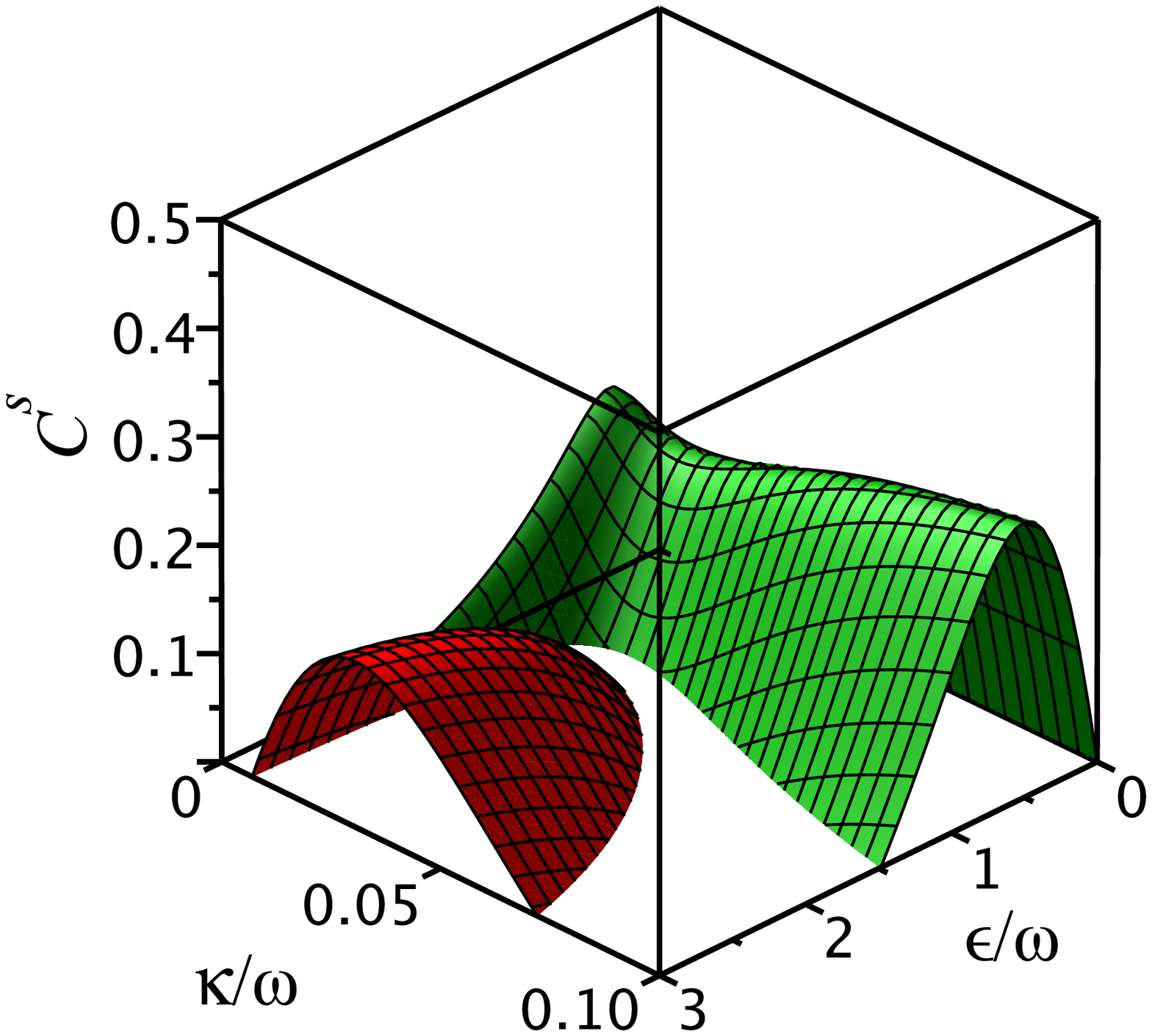}
\caption{(Color online) Stationary concurrence $\mathcal{C}^{s}$ as a function of the coupling strengths $\kappa$ and $\epsilon$ when the cavity modes decay to separate reservoirs, $\gamma =0$. The damping rate $\gamma_B$ is fixed at $\gamma_B = 0.01\omega$ and (a) $\gamma_A =0.01\omega$, (b) $\gamma_A =0.1\omega$ and (c) $\gamma_A =0.2\omega$. The red surface represents a contribution of $C_{1}^{s}$, while the green (light gray) part represents the contribution of $C_{2}^{s}$ to the entanglement created between the modes.}
\label{fig3}
\end{figure}

The concurrence given by Eq.~(\ref{20}) is plotted in Fig.~\ref{fig3} as a function of the coupling strengths $\kappa$ and $\epsilon$. For the balanced decay, Fig.~\ref{fig3}(a), the entanglement is independent of $\kappa$ and occurs in a range of $\epsilon<\sqrt{\gamma_{0}^2+4\omega^2}$. For the unbalanced decay, Figs.~\ref{fig3}(b) and (c), there are two separate ranges of the parameters where entanglement occurs. As discussed above, these two ranges are determined by $C_{1}^{s}>0$ and $C_{2}^{s}>0$, respectively. We see a gap between the $C_{2}^{s}>0$ and $C_{1}^{s}>0$ structures that entanglement created by the one- and two-photon coherences lies in separate ranges of the parameters.
 Moreover, the magnitude of $C_{2}^{s}$ is reduced in the range of $\kappa$ where $C_{1}^{s}$ emerges. Evidently, with an increasing asymmetry between the damping rates the entanglement shifts from $C_{2}^{s}$ to $C_{1}^{s}$.  Thus, the creation of entanglement by the coherence $\rho_{23}$ occurs in expense of the entanglement created by the coherence~$\rho_{14}$. One can also notice from Fig.~\ref{fig3} that the entanglement as determined by $C_{2}^{s}$ occurs in the parameters range $\kappa/\omega \ll1$ and $\epsilon/\omega \approx 1$. On the other hand, the entanglement as determined by $C_{1}^{s}$ occurs in the deep strong coupling regime of $\epsilon/\omega >1$.

It is interesting to examine which of the two quantities, $C_{1}^{s}$ or $C_{2}^{s}$, produces the largest degree of entanglement and whether the maximum corresponds to the case of distinguishable or indistinguishable modes. A quick inspection of Eq.~(\ref{22}) shows that $C_{2}^{s}$ achieves its maximum value at $\gamma_{d}=0$ and the corresponding maximum value is
\begin{equation}
C_{2}^{s} = \frac{2\epsilon\left(\sqrt{4\omega^2+\gamma_{0}^2} -\epsilon\right)}{4\omega^2+4\epsilon^{2}+\gamma_{0}^2} .
\end{equation}
Viewed as a function of $\epsilon$, $C_{2}^{s}$ is maximal at $\epsilon = \sqrt{4\omega^2+\gamma_{0}^2}/4$, in which case $C_{2}^{s}=3/10$.

An inspection of Eq.~(\ref{21}) reveals that the maximum value of $C_{1}^{s}$ occurs for $|\gamma_{d}|=\gamma_{0}$, corresponding to either $\gamma_{A}\gg\gamma_{B}$ or $\gamma_{B}\gg\gamma_{A}$, and when $\epsilon,\gamma_{0}\gg \kappa$. In these limits,  $C_{1}^{s}$ is small, $C_{1}^{s} = 2\kappa/\gamma_{0}\ll 1$.
This means that the largest degree of entanglement produced in the system is that determined by $C_{2}^{s}$. Since for $|\gamma_{d}|= 0$, at which $C_{2}^{s}$ attains its maximum value, the visibility ${\cal V}_{s}=0$, we conclude the largest degree of entanglement is achieved when the modes are completely distinguishable.

\subsection{The case of collective decay, $\gamma\neq 0$}

Let now turn to the case of the decay of the modes to a common reservoir with $\gamma=\sqrt{\gamma_{A}\gamma_{B}}$. The corresponding steady-state solution for the density matrix elements are given by Eq.~(\ref{24a}). When applying Eq.~(\ref{24a}) to the concurrence, we find the following expressions for the quantities $C_{1}^{s}$ and $C_{2}^{s}$:
\begin{eqnarray}\label{25b}
C_{1}^{s} &=& \frac{2\epsilon}{\tilde{D}}\Bigg[\frac{1}{2}\epsilon\sqrt{\kappa^{2}(\gamma_{A}-\gamma_{B})^{2} +\gamma_{A}\gamma_{B}\gamma_{0}^{2}} \nonumber\\
&&-\kappa^{2}\sqrt{\epsilon^2+4\omega^2+\gamma_{0}^2}\Bigg] ,
\end{eqnarray}
and
\begin{align}\label{25c}
  C_{2}^{s} &= \frac{2\epsilon}{\tilde{D}}\Bigg[\kappa^{2}\sqrt{\gamma_{0}^{2} + 4\omega^2} \nonumber\\
&-\frac{1}{2}\epsilon\sqrt{(2\kappa^2+\gamma_{A}\gamma_{0})(2\kappa^2+\gamma_{B}\gamma_{0})}\Bigg] .
\end{align}
We see that in the case of damping of the modes to a common reservoir, the role of $\kappa$ in the creation of entanglement reversed, a nonzero $\kappa$ is now required for $C_{2}^{s}$ to be positive, whereas $C_{1}^{s}$ can be positive even for $\kappa =0$.

If we set $\kappa=0$ in Eqs.~(\ref{25b}) and (\ref{25c}), we find $C_{2}^{s}<0$ and $C_{1}^{s}=\sqrt{\gamma_{A}\gamma_{B}}/\gamma_{0}$. In this case, the concurrence is insensitive to $\epsilon$. Therefore, the modes can be entangled for all values
of~$\epsilon$. The reason is that now the mechanism responsible for the generation of the entanglement is in the trapping of the population in the state $\ket d$. Note that the concurrence depends only on the damping rates and therefore can be close to the optimum value of unity, which can be achieved for $\gamma_{A}\approx \gamma_{B}$.

\begin{figure}[h]
\hspace*{-6.3cm}\textbf{(a) } \\ \textbf{(b)}\includegraphics[width=.72\columnwidth]{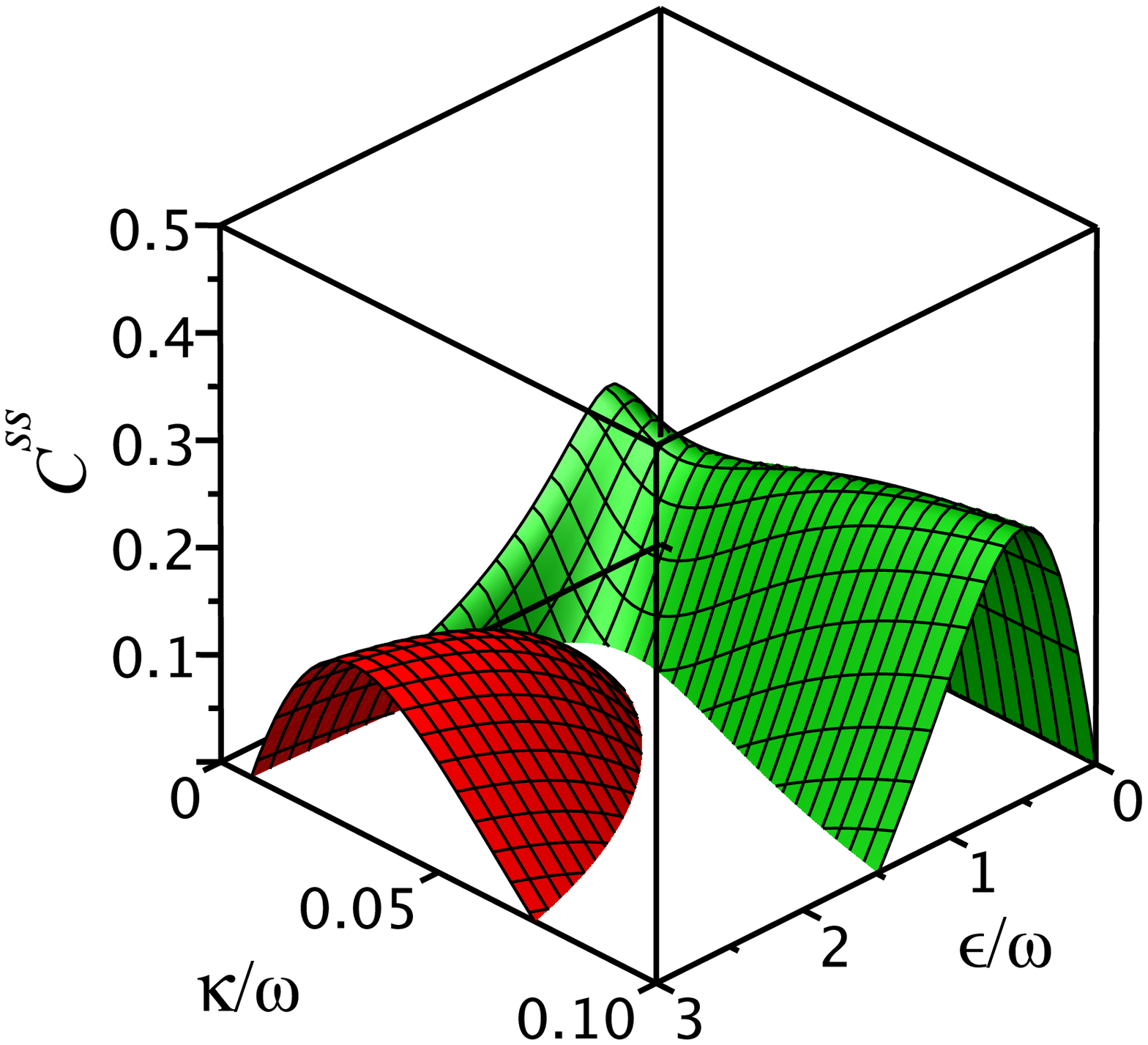}\\
\textbf{(c)}\includegraphics[width=.72\columnwidth]{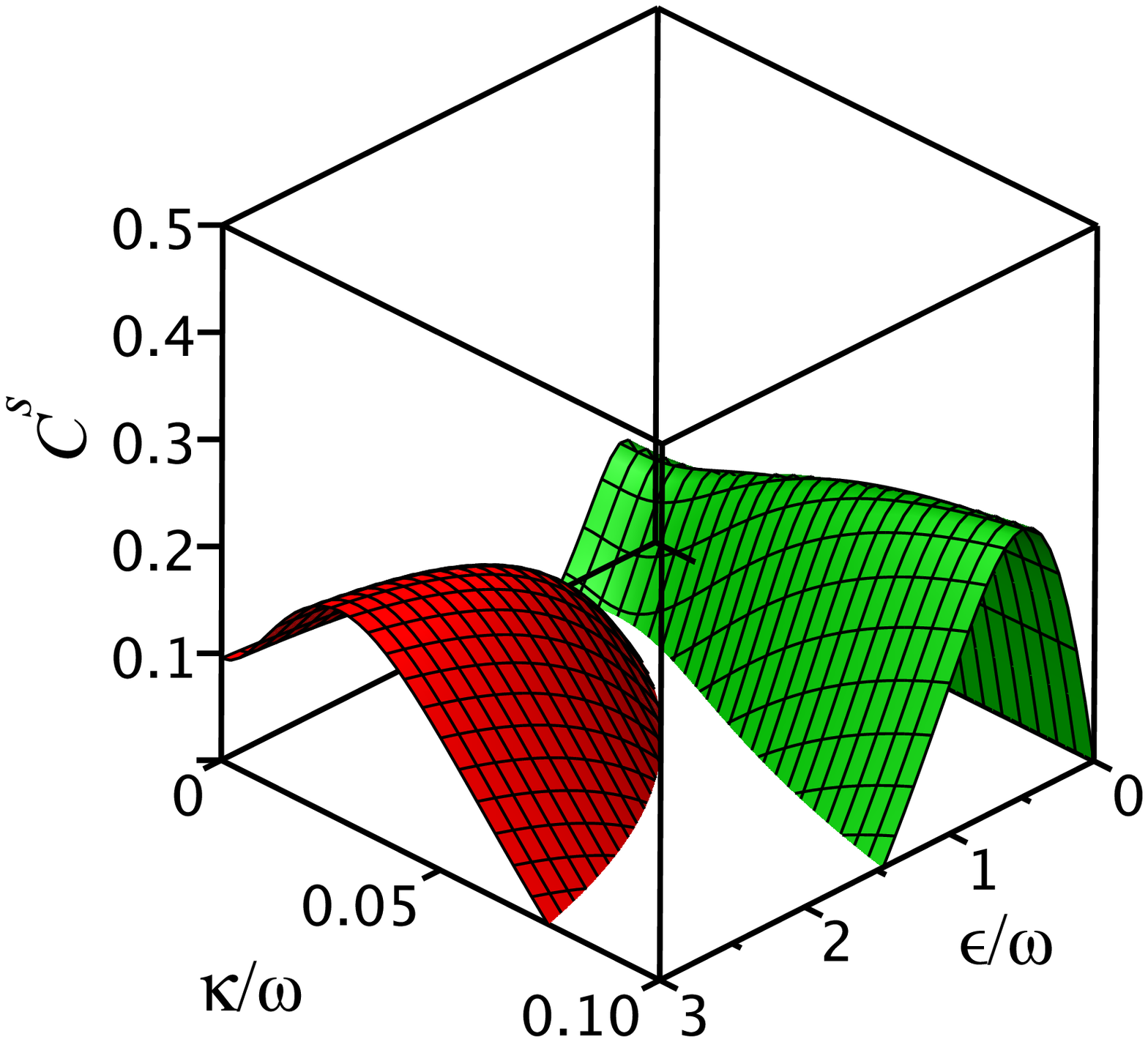}\\
\includegraphics[width=.72\columnwidth]{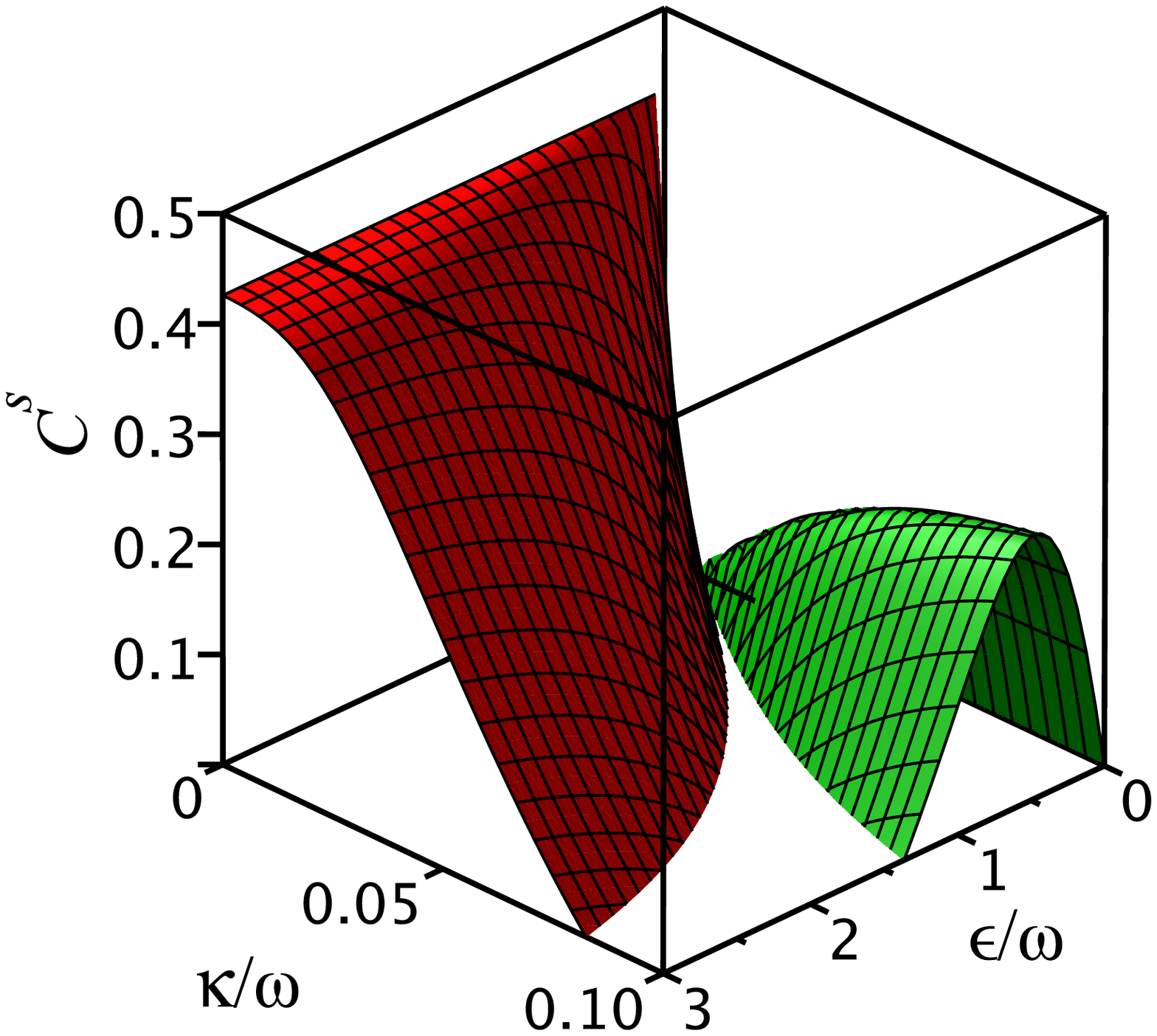}
\caption{(Color online) Stationary concurrence $\mathcal{C}^{s}$ as a function of the coupling strengths $\kappa$ and $\epsilon$ when the cavity modes decay to a common reservoir. The plots are for $\gamma_A =0.2\omega$, $\gamma_B = 0.01\omega$ and different $\gamma$: (a) $\gamma = 0$, (b) $\gamma = \frac{1}{2}\sqrt{\gamma_{A}\gamma_{B}}$ and (c) $\gamma = \sqrt{\gamma_{A}\gamma_{B}}$. The red surface represents the contribution of $C_{1}^{s}$, while the right green (light gray) surface represents the contribution of $C_{2}^{s}$.}
\label{fig4}
\end{figure}

Figure~\ref{fig4} shows the effect of the collective damping $\gamma$ on the concurrence of the modes. We see that the collective decay results in an entanglement which is associated mostly with the quantity $C_{1}^{s}$. Hence, it is mostly associated with the presence of the coherence $\rho_{23}$. Moreover, the concurrence although the most positive in the strong coupling regime of the antiresonant terms, it seen to be positive in the weak coupling regime of both resonant $(\kappa/\omega \ll 1)$ and antiresonant $(\epsilon/\omega \ll 1)$ terms.

Comparing $C_{1}^{s}$ with the visibility, Eq.~(\ref{b42}), we easily find that in the case of $\kappa =0$, where the system evolves to the pure state $\ket d$, $C_{1}^{s}$ is equal to ${\cal V}_{c}$. Thus, in the case of pure states there is a direct connection between indistinguishability and entanglement~\cite{jb03,sm10}. Otherwise, when the system is in a mixed state, one can observe entanglement with the complete distinguishability of the modes and vice versa.

\begin{figure}[h]
  \includegraphics[width=0.8\columnwidth]{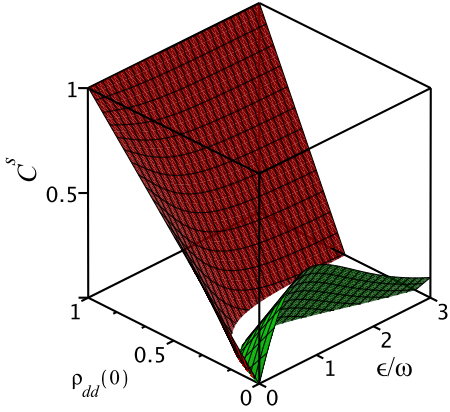}
  \caption{(Color online) Stationary concurrence in terms of $\epsilon$ and the initial condition $\rho_{dd}(0)$ for $\gamma_a=\gamma_B=\gamma=0.01\omega$.}
  \label{fig5}
\end{figure}

Finally, we consider the case of the balanced decay of the modes $(\gamma_{A}=\gamma_{B}\equiv \gamma_{0})$ with the collective damping rate $\gamma =\gamma_{0}$. We have shown in Sec.~\ref{sec3b} that in this special case the steady-state values of the density matrix elements depend on initial conditions. Moreover, it is independent of $\kappa$. Therefore, it is straightforward, using the results given in Eq.~(\ref{26b}), to show that entanglement between the modes is related to the initial state. Specifically, if initially the system is prepared in the maximally entangled state $\ket d$, it will remain in this state for all times. If the initial state is different for $\ket d$ then the system can decay to an entangled state created by $\epsilon$. This is illustrated in Fig.~\ref{fig5}, where we plot the concurrence as a function of $\epsilon$ and the initial population of the state $\ket d$.

Similarly to the case of the unbalanced decay presented in Fig.~\ref{fig3}, the entanglement is mostly associated with the presence of the coherence $\rho_{23}$. Only for initial states at which $\rho_{dd}(0)\approx 0$, the entanglement created is associated with the coherence $\rho_{14}$. Moreover, the entanglement, which is independent of the coupling strength $\kappa$ of the resonant terms, is present in all ranges of the coupling strength $\epsilon$ of the antiresonant terms.

Comparing the concurrence with the visibility, we see that in the case of the collective decay of the modes, the maximum entanglement is achieved when the modes are indistinguishable, and the maximum possible entanglement of $C^{s}=1$ is achieved when the modes are completely indistinguishable, ${\cal V}=1$. Therefore, we may conclude that in the case the collective decay of the modes, more entanglement is achieved with more indistinguishability and the maximum possible entanglement is achieved with completely indistinguishable modes.

\subsection{Second-order correlations}

We have seen that the creation of entanglement is determined by two criteria $C_{1}^{s}$ and $C_{2}^{s}$ which do not overlap. In other words, these two criteria determine two distinct ranges of the parameters at which entanglement occurs. We may relate these criteria to the normalized second-order photon-photon correlation function $g^{(2)}(0)$ which is directly measurable in coincidence counting schemes and provides a test of whether the photons are correlated (bunched) or anti-correlated (antibunched)~\cite{zz16}. For this purpose, we consider the normalized second-order correlation function, which for the two modes $A$ and $B$ is~\cite{pa82}
\begin{align}
g^{(2)}(0) \equiv \frac{\langle\hat{a}_{A}^{\dag}\hat{a}_{B}^{\dag}\hat{a}_{A}\hat{a}_{B}\rangle}{\langle\hat{a}_{A}^{\dag}\hat{a}_{A}\rangle\langle \hat{a}_{B}^{\dag}\hat{a}_{B}\rangle} =\frac{\rho_{44}}{(\rho_{44}+\rho_{22})(\rho_{44}+\rho_{33})} .
\end{align}
If we compare this expression with the expressions for $C_{1}^{s}$ and $C_{2}^{s}$, Eqs.~(\ref{11}) and (\ref{12}), we find that there is no direct connection here between entanglement and the second-order photon correlations. The quantities $C_{1}^{s}$ and $C_{2}^{s}$ are given in terms of the coherences and populations, while $g^{(2)}(0)$ is given entirely in term of the populations.
Nevertheless, we can demonstrate that entanglement determined by $C_{1}^{s}>0$ occurs in the range of the parameters at which $g^{(2)}(0)<1$, whereas entanglement determined by $C_{2}^{s}>0$ occurs in the range at which $g^{(2)}(0)>1$.

Let us examine the relations for the case of independent reservoirs $(\gamma =0)$. To do this, let us first assume that $\gamma_{A}=\gamma_{B}$. Then using Eqs.~(\ref{18}) and (\ref{19}) we readily find
\begin{align}
g^{(2)}(0) = 1 + \frac{4\omega^{2}+\gamma_{0}^{2}}{4\epsilon^{2}} ,
\end{align}
from which it is clear that $g^{(2)}(0)$ is always greater than one. This means that emitted photons exhibit bunching effect when $\gamma_{A}=\gamma_{B}$.

On the other hand, when $\gamma_{A}\neq \gamma_{B}$ and in the limit of $\kappa< \epsilon$, it can be shown that $g^{(2)}(0)$ is of the form
\begin{align}
g^{(2)}(0)\approx 1 - \frac{4\kappa^{2}\gamma_{d}^{2}}{(4\kappa^{2}+\gamma_{0}^{2})^{2}-\gamma_{0}^{2}\gamma_{d}^{2}} .
\end{align}
Here we see that $g^{(2)}(0)$ is always less than one. It follows that for $\gamma_{A}\neq \gamma_{B}$ and $\kappa <\epsilon$, the emitted photons exhibit antibunching effect.
\begin{figure}[h]
\hspace*{-6.3cm}\textbf{(a) } \\ \textbf{(b)}\includegraphics[width=.72\columnwidth]{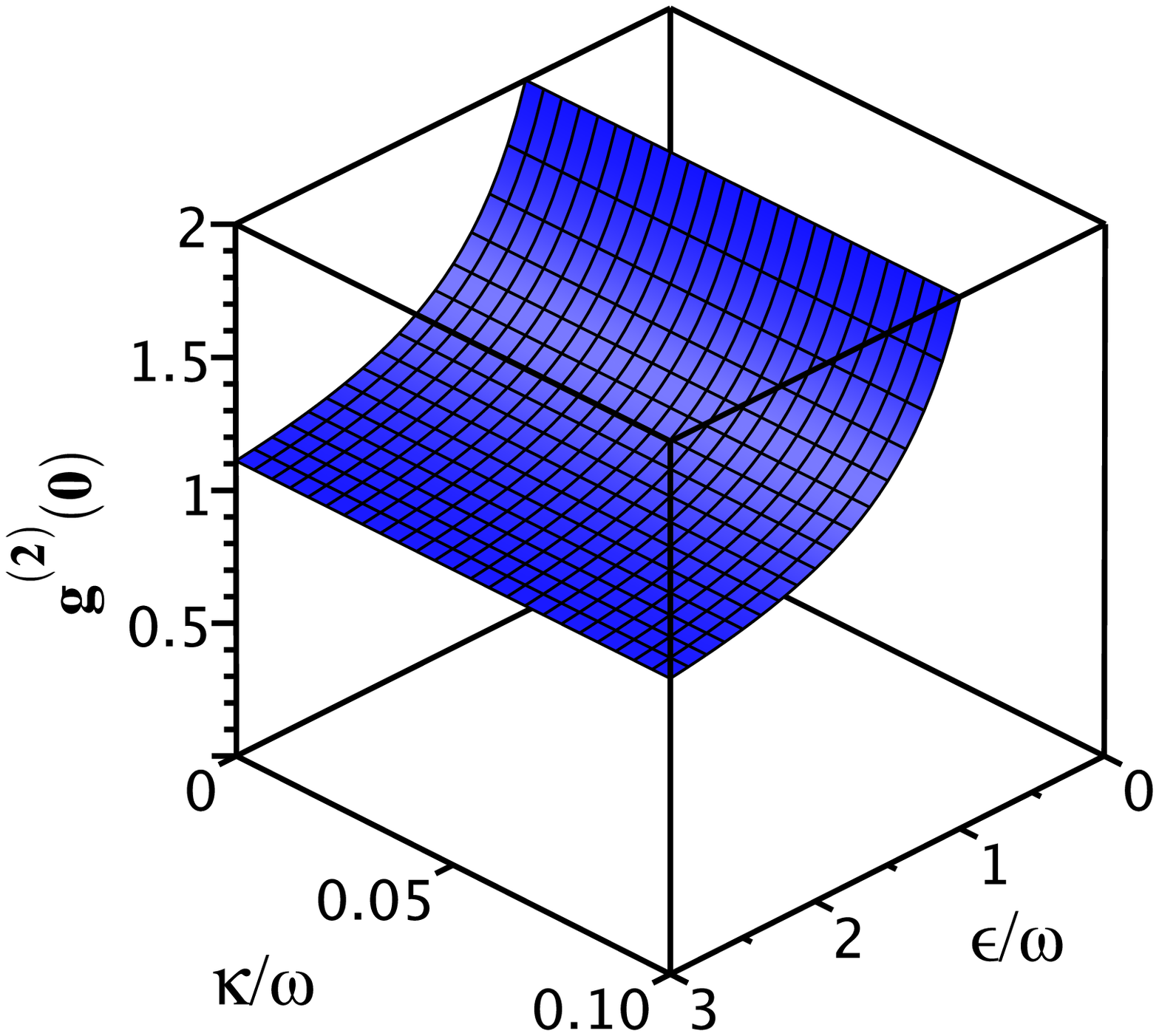}\\
\textbf{(c)}\includegraphics[width=.72\columnwidth]{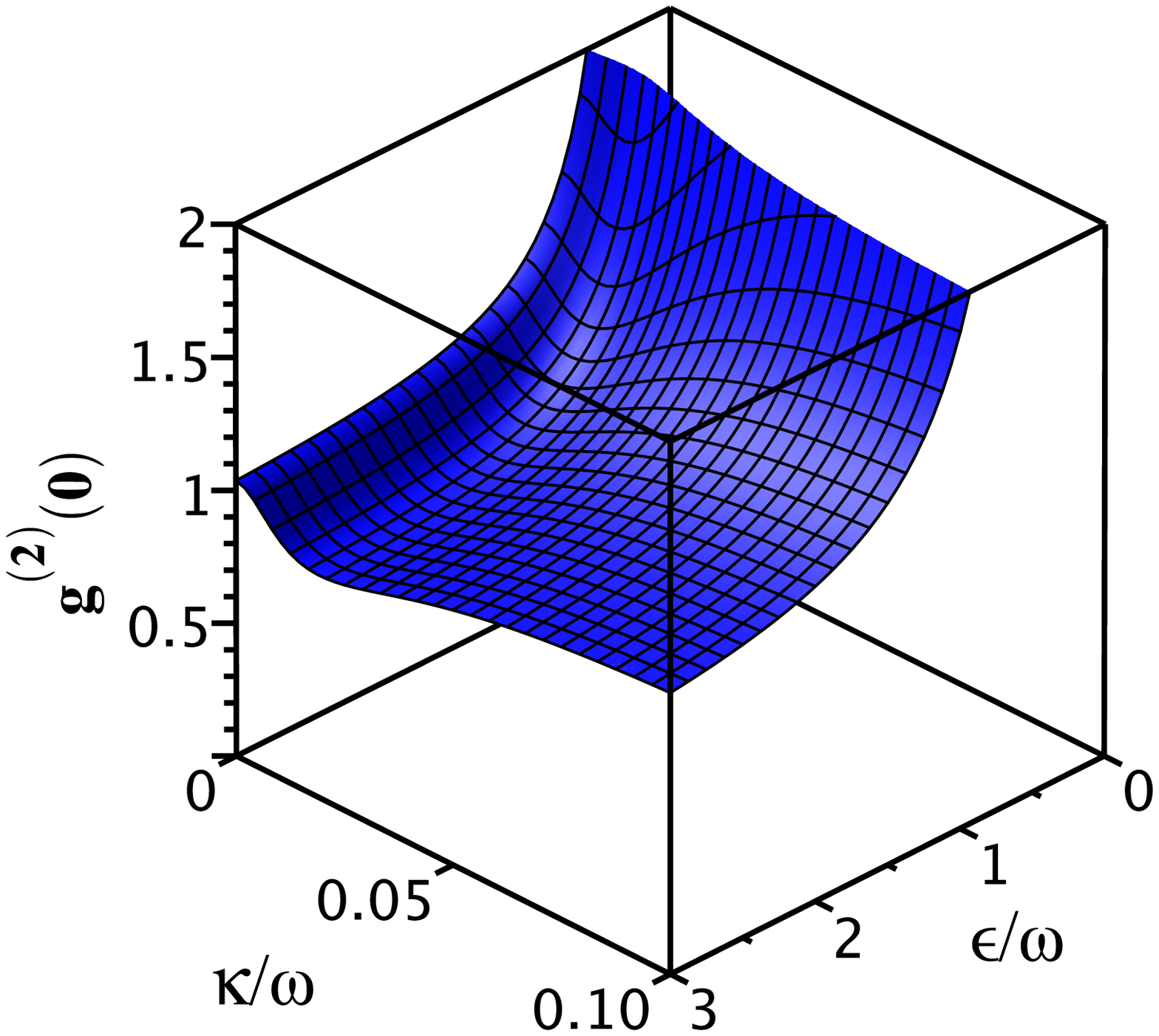}\\
\includegraphics[width=.72\columnwidth]{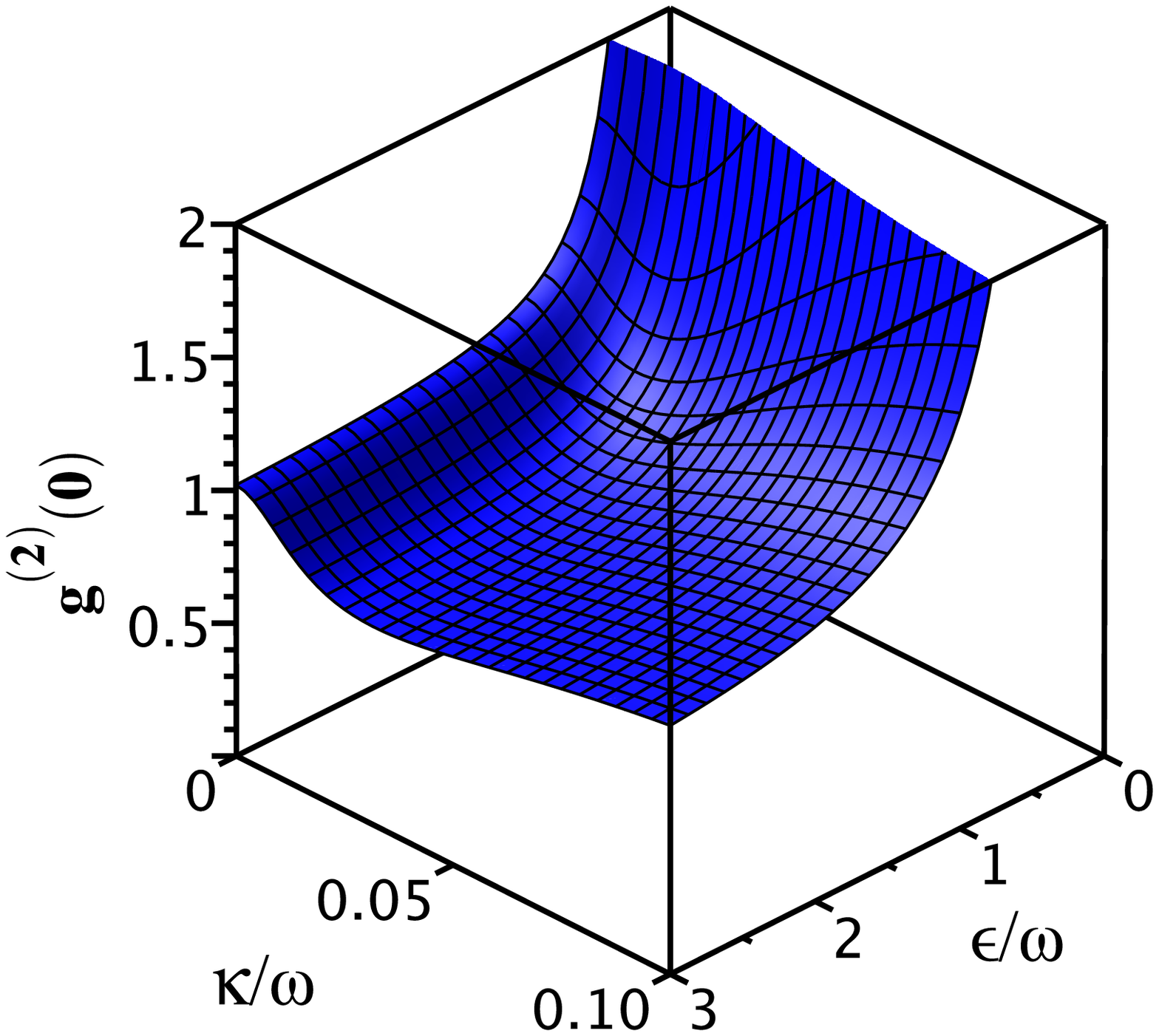}
\caption{(Color online) Stationary second-order correlation function $g^{(2)}(0)$ plotted as a function of the coupling strengths $\kappa$ and $\epsilon$ for the same parameters as in Fig.~\ref{fig3}.}
\label{fig6}
\end{figure}

Figure~\ref{fig6} shows the variation of $g^{(2)}(0)$ with $\kappa$ and $\epsilon$ for the same parameters as in Fig.~\ref{fig3}, where we illustrated the variation of ${\cal C}^{s}$ with $\kappa$ and $\epsilon$.
It is seen that $g^{(2)}(0)$ decreases with an increasing $\epsilon$ and for $\gamma_{A}=\gamma_{B}$ attains a minimum value of $g^{(2)}(0)=1$ independent of $\kappa$. For $\gamma_{A}\neq \gamma_{B}$, there is a range of $\kappa\, (\kappa\ll\epsilon)$ at which $g^{(2)}(0)<1$. Comparing Fig.~\ref{fig6} with Fig.~\ref{fig3}, we see that the positive values of $C_{2}^{s}$ lie within the parameters ranges permissible for photon bunching, $g^{(2)}(0)>1$, and the positive values of $C_{1}^{s}$ lie within the permissible ranges for photon antibunching, $g^{(2)}(0)<1$.

Summarizing, there is a connections between entanglement and photon statistics that the entanglement determined by $C_{1}^{s}>0$ is related to photon antibunching whereas the entanglement determined by $C_{2}^{s}>0$ is related to photon bunching effect.

\section{Summary and conclusions}\label{sec6}

We have investigated two concepts of quantum mechanics, indistinguishability and entanglement, in a system composed of two strongly coupled bosonic modes. We have found that the use of both resonant (RWA) and antiresonant (non-RWA) terms in the interaction between the modes forms a natural link of the two concepts with the energy-time uncertainty principle. The inclusion of the antiresonant terms requires to work in an ultra-strong coupling regime and at a very short interaction time. We have found nonzero population distribution and coherences between the low energy states and have interpreted the distribution as the result of the uncertainty in energy which over a very short interaction time is of the order of the one-photon energy.

The analysis of the steady-state of the system has demonstrated the importance of the dissipation in the redistribution of the population and in the creation of coherences between the low energy states. To explore the role of the dissipation, we have calculated the steady-state of the system when the modes decay either independently or collectively. We have found that when the modes decay independently, the distinguishability and entanglement of the modes depend strongly on whether the modes decay with equal or unequal rates. In particular, when the modes decay with equal rates, entanglement with the complete distinguishability of the modes can be observed; the entangled cavity modes behave as mutually incoherent. When the modes decay with unequal rates, a single-photon coherence is induced between the modes resulting in indistinguishability, single-photon interference between the modes. We have found an upper bound of the single-photon visibility that the visibility cannot exceed $50\%$ when the modes decay independently. When the modes decay with equal rates we show that "which-path" information is made possible and the visibility in single-photon interference vanishes.

When the modes decay collectively, the single-photon coherence is created even if the modes decay with equal rates. The additional pathway induced by the collective decay rate results in nearly perfect visibility of the interference pattern even in the absence of the resonant coupling between the modes. We have shown that the collective damping creates superposition states and in the absence of the resonant coupling the steady-state is a pure entangled state rather than the expected mixed state. This can result in entanglement with the complete indistinguishability of the modes, that the modes entangled through the collective decay behave as mutually coherent. In addition, we have found that the collective damping with the maximal rate can lead to the steady-state values for the density matrix elements which depend on their initial values. This requires that the modes decay with equal rates. Then, depending on the initial state, the modes can be found in a mixed or in a pure maximally entangled state.

\end{document}